\documentclass[sigconf]{acmart}

\usepackage{xcolor}
\usepackage{enumitem}
\usepackage{amsmath}
\usepackage{eucal}

\usepackage{xcolor}
\usepackage[linesnumbered,ruled,lined]{algorithm2e}

\SetCommentSty{mycommfont}

\newcommand{\so}[0]{Stack Overflow}

\newcommand{\toolname}[0]{\textsc{TechSumBot}}
\newcommand{\firstmodule}[0]{Usefulness Ranking}
\newcommand{\secondmodule}[0]{Centrality Estimation}
\newcommand{\thirdmodule}[0]{Redundancy Removal}

\newcommand{\nlp}[0]{NLP community}

\newcommand{\bench}[0]{\textsc{TechSumBench}}

\newif\ifunnumfnt

\copyrightyear{2022} 
\acmYear{2022} 
\setcopyright{acmlicensed}\acmConference[ASE '22]{37th IEEE/ACM International Conference on Automated Software Engineering}{October 10--14, 2022}{Rochester, MI, USA}
\acmBooktitle{37th IEEE/ACM International Conference on Automated Software Engineering (ASE '22), October 10--14, 2022, Rochester, MI, USA}
\acmPrice{15.00}
\acmDOI{10.1145/3551349.3560421}
\acmISBN{978-1-4503-9475-8/22/10}

\begin{document}

\title{Answer Summarization for Technical Queries: \\Benchmark and New Approach}

\author{Chengran Yang\textsuperscript{$\dag$}, Bowen Xu\textsuperscript{$\dag$}\textsuperscript{$\ast$}, Ferdian Thung\textsuperscript{$\dag$}, Yucen Shi\textsuperscript{$\dag$}, Ting Zhang\textsuperscript{$\dag$}, Zhou Yang\textsuperscript{$\dag$}, Xin Zhou\textsuperscript{$\dag$}, Jieke Shi\textsuperscript{$\dag$}, Junda He\textsuperscript{$\dag$}, DongGyun Han\textsuperscript{$\ddag$}, David Lo\textsuperscript{$\dag$}}

\affiliation{%
  \institution{\textsuperscript{$\dag$}School of Computing and Information Systems, Singapore Management University}
  \country{}
}
\email{{cryang,bowenxu.2017,ferdianthung,ycshi,tingzhang.2019,zyang,xinzhou.2020,jiekeshi,jundahe,davidlo}@smu.edu.sg}
\affiliation{%
  \institution{\textsuperscript{$\ddag$}Royal Holloway, University of London}
  \country{}
}
\email{donggyun.han@rhul.ac.uk}

\renewcommand{\shortauthors}{Yang et al.}

\begin{abstract}
Prior\unnumfnttrue\footnotetext{$\ast$ Corresponding author} studies have demonstrated that approaches to generate an answer summary for a given technical query in Software Question and Answer (SQA) sites are desired.
We find that existing approaches are assessed solely through user studies. Hence, a new user study needs to be performed every time a new approach is introduced; this is time-consuming, slows down the development of the new approach, and results from different user studies may not be comparable to each other.
There is a need for a benchmark with ground truth summaries as a complement assessment through user studies. 
Unfortunately, such a benchmark is non-existent for answer summarization for technical queries from SQA sites.

To fill the gap, we manually construct a high-quality benchmark to enable automatic evaluation of answer summarization for the technical queries for SQA sites. 
It contains 111 query-summary pairs extracted from 382 Stack Overflow answers with 2,014 sentence candidates. 
Using the benchmark, we comprehensively evaluate the performance of existing approaches and find that there is still a big room for improvements.

Motivated by the results, we propose a new approach \toolname{} with three key modules:
1) \textit{Usefulness Ranking} module; 2) \textit{Centrality Estimation} module; and 3) \textit{Redundancy Removal} module.
We evaluate \toolname{} in both automatic (i.e., using our benchmark) and manual (i.e., via a user study) manners. 
The results from both evaluations consistently demonstrate that \toolname{} outperforms the best performing baseline approaches from both SE and NLP domains by a large margin, i.e., 10.83\%--14.90\%, 32.75\%--36.59\%, and 12.61\%--17.54\%, in terms of ROUGE-1, ROUGE-2, and ROUGE-L on automatic evaluation, and 5.79\%--9.23\% and 17.03\%--17.68\%, in terms of average usefulness and diversity score on human evaluation.
This highlights that automatic evaluation  on our benchmark can uncover findings similar to the ones found through user studies. More importantly, the automatic evaluation has a much lower cost, especially when it is used to assess a new approach.
Additionally, we also conducted an ablation study, which demonstrates that each module in \toolname{} contributes to boosting the overall performance of \toolname{}.
We release the benchmark as well as the replication package of our experiment at \url{https://github.com/TechSumBot/TechSumBot}.

\end{abstract}

\unnumfntfalse


\maketitle

\section{Introduction}
Online software question and answer (SQA) forums are becoming an integral part of software engineering. As the Stack Overflow community flourishes, the platform has around 23 million existing answers as of May 2022, making it a valuable software engineering resource.\footnote{\url{https://stackexchange.com/sites?view=list\#traffic}} Developers often rely on Stack Overflow to acquire software knowledge, such as learning API usage, fixing bugs, and discovering trends in development technologies~\cite{uddin2017automatic,silva2019recommending,yang2022aspect,xu2017answerbot,nadi2020essential,mahajan2020recommending} by searching from a massive collection of existing answers. Unfortunately, multiple works (e.g.,~\cite{xu2017answerbot,liu2021characterizing}) have demonstrated that developers suffer from the ineffective and time-consuming answer searching process for their technical queries due to useless, redundant, and incomplete information returned by existing search engines.


Several studies have been proposed to help developers capture desired online information more accurately and succinctly by proposing query-focused summarization approaches \cite{nadi2020essential, treude2016augmenting,wang2019extracting,xu2017answerbot, silva2019recommending} -- this task is often referred to as {\em answer summarization for technical questions}.
The closest work is proposed by Xu et al. \cite{xu2017answerbot}, who proposed the latest Stack Overflow answer summarization approach named AnswerBot for technical queries. The problem is formulated as a query-focused extractive summarization problem, 
i.e., selecting an optimal subset of $k$ sentences from multiple answers to form a summary to answer the target technical query.

However, we find that the work carries a major limitation on the evaluation. Both AnswerBot and its considered baseline approaches are evaluated only in a manual manner through user studies. Although user studies can simulate the realistic usage environment of the approaches, it carries two drawbacks at the same time. First, user studies are usually expensive in terms of time cost and human resources. 
It requires a long period of time for preparation, labeling, and human feedback collection.
Second, the user study needs to be repeated every time a new approach is introduced. Aside from slowing down the development of new approaches, it also raises fairness issues. To ensure a fair comparison, a user study on all the considered approaches requires to be done under a similar evaluation setting. However, this fairness constraint is hard to enforce for a series of user studies that evaluate different newly proposed approaches. Such user studies are often performed with different groups of participants and different sets of technical questions.

To mitigate the aforementioned limitations of user studies, automatic evaluation through a benchmark is in need as it can be a good complement.
After the initial investment to set up a benchmark, future studies can reuse it to evaluate new approaches in an identical evaluation setting to produce easily repeatable results.
For the answer summarization for technical questions task, a benchmark ideally consists of a set of triplets in which each triplet contains a technical query $Q$, a set of answers $Ans$, and ground truth answer summaries $S$ with respect to the query, i.e., $Bench=\{\langle Q, S, Ans\rangle\}$.
Such a benchmark is often constructed for various summarization tasks in the Natural Language Processing (NLP) field, e.g., CNN/Daily Mail\cite{nallapati2016abstractive}, GigaWord~\cite{napoles2012annotated} and DUC--2005\cite{dang2005overview}. However, no such benchmark exists for our task considering SQA sites.

Motivated by this, we conduct the first Stack Overflow query-focused multi-answer summarization benchmark \bench{} (described in Section~\ref{sec:benchmark}). 
Table \ref{tab:example summary} demonstrates an example pair of a technical query and its corresponding answer summary. 
\vspace{-5mm}
\begin{table}[htbp]
    \centering
    \caption{Example answer summary in \bench{} with respect to a technical query.}
    \vspace{-3mm}
    \begin{tabular}{|p{8cm}|}\hline
\textbf{Technical Query: Difference between Spring MVC and Spring Boot} \\\hline
\textbf{Ground Truth Summary in \bench{}:}\\
1. $\langle$ Spring MVC is a sub-project of the Spring Framework, targeting design and development of applications that use the MVC (Model-View-Controller) pattern.$\rangle$ \\
2. $\langle$  Spring boot is a utility for setting up applications quickly, offering an out of the box configuration in order to build Spring-powered applications.$\rangle$\\
3. $\langle$ Spring boot = Spring MVC + Auto Configuration(Don't need to write spring.xml file for configurations) + Server(You can have embedded Tomcat, Netty, Jetty server).$\rangle$ \\
4. $\langle$ Spring MVC framework is module of spring which provide facility HTTP oriented web application development.$\rangle$ \\
5. $\langle$ So, Spring MVC is a framework to be used in web applications and Spring Boot is a Spring based production-ready project initializer.$\rangle$   \\
\hline
    \end{tabular}
    \label{tab:example summary}
\vspace{-6mm}

\end{table}

\bench{} enables us to fairly compare query-focused summarization approaches proposed in the software engineering and NLP fields in an identical evaluation setting and in a repeatable manner.
We have used \bench{} to investigate four approaches (i.e., AnswerBot~\cite{xu2017answerbot}, LexRank~\cite{erkan2004lexrank}, Biased-TextRank~\cite{kazemi2020biased}, QuerySum~\cite{xu2020coarse}) and find that QuerySum performs the best, but there is still a big room for improvement. Furthermore, we performed a qualitative analysis to seek the reason for the poor performance of these approaches (described in Section~\ref{sec:qualitative}) and summarized two key findings. First, the handcrafted features considered in the existing approaches have limitations. For example, one handcrafted feature in AnswerBot (e.g., \emph{semantic patterns}) uses 12 patterns to indicate the importance of answer sentences, e.g., check if a sentence contains the string ``I suggest that...''. Such a semantic pattern is not versatile enough to capture the semantics of questions and answer sentences.
Second, when removing redundancy from a set of sentences, we find that sentence representation plays an important role in determining the semantic relatedness between sentences.

To better tackle the problem, we propose \toolname{} (described in Section~\ref{sec:method}), a new query-focused answer summarization approach with three core modules: 1) \firstmodule{}; 2) \secondmodule{}; 3) \thirdmodule{}.
Correspondingly, they take into consideration 1) the usefulness of each answer sentence with respect to the query, 2) sentence centrality among all candidates, and 3) semantic similarity between sentences. 
In the \firstmodule{} module, we rank the \textit{usefulness} of each sentence to the query by leveraging a transfer learning-based pre-trained Transformer model. 
In the \secondmodule{} module, we re-rank the sentence by estimating the central sentences among all candidates.
Finally, in the \thirdmodule{} module, we remove redundant sentences by applying a greedy selection mechanism, in which we leverage a SE domain sentence representation model to calculate the similarity between sentences.
Particularly, as no existing dataset in SE domain can be directly used for supervising the sentence representation model, we carefully observe the characteristic of Stack Overflow data and create an in-domain sentence relationship dataset by leveraging the semantics inferred from duplicate question links and tags information; both of them are manually labeled by Stack Overflow users in a high-standard mechanism.
Then we use our in-domain sentence relationship dataset to enhance the state-of-the-art contrastive learning-based sentence representation model SIMCSE~\cite{gao2021simcse}. 


To facilitate a comprehensive evaluation of \toolname, we conduct an automatic evaluation using TechSumBench and a user study.
The automatic evaluation results show that \toolname{} substantially outperforms the best performing baselines in both SE and NLP fields by 10.83\%--14.90\%, 32.75\%--36.59\%, and 12.61\%--17.54\% in terms of ROUGE-1, ROUGE-2, and ROUGE-L, respectively.
The result of the user study is consistent with the result of the automatic evaluation: TechSumBot substantially outperforms the best performing baseline
by 5.79\%--9.23\%, 17.03\%--17.68\% in terms of average usefulness and diversity scores. Besides, an ablation study is performed and its result demonstrates that each module complementarily contributes to the overall performance.

The main contributions of this paper are the following:
\begin{itemize}[nosep,leftmargin=*]
	\item We construct the first summarization benchmark named \bench{} for answer summarization of technical queries in SQA sites. The benchmark can be used to automatically evaluate future proposed summarization methods in an identical setting and a repeatable manner.
	\item We comprehensively evaluate four query-focused summarization approaches from both SE and NLP fields using our benchmark. The experiments highlight the limitations of the approaches and indicate a room for improvement.
	\item We propose \toolname{}, a novel query-focused answer summarization approach with three modules to better solve the problem.
	\item We evaluate the performance of \toolname{} via both automatic evaluation and user study. The results of both evaluations demonstrate that \toolname{} outperforms the best performing baselines by a large margin.
	\item We release the benchmark, as well as the replication package of \toolname{} to facilitate future work.

\end{itemize}

The rest of this paper is structured as follows. Section \ref{benchmark} describes the benchmark construction. Section 3 presents the framework of our proposed \toolname{}. Section 4 introduces the baselines and metrics for automatic evaluation. Section 5 analyzes the experiment result. Section 6 discusses the qualitative analysis and threats to validity. Section 7 surveys the related work. Section 8 concludes our work and presents our future plan.

\section{Benchmark}\label{sec:benchmark}
\label{benchmark}

\begin{figure}[htbp]
    \centering
    \includegraphics[width=8.5cm]{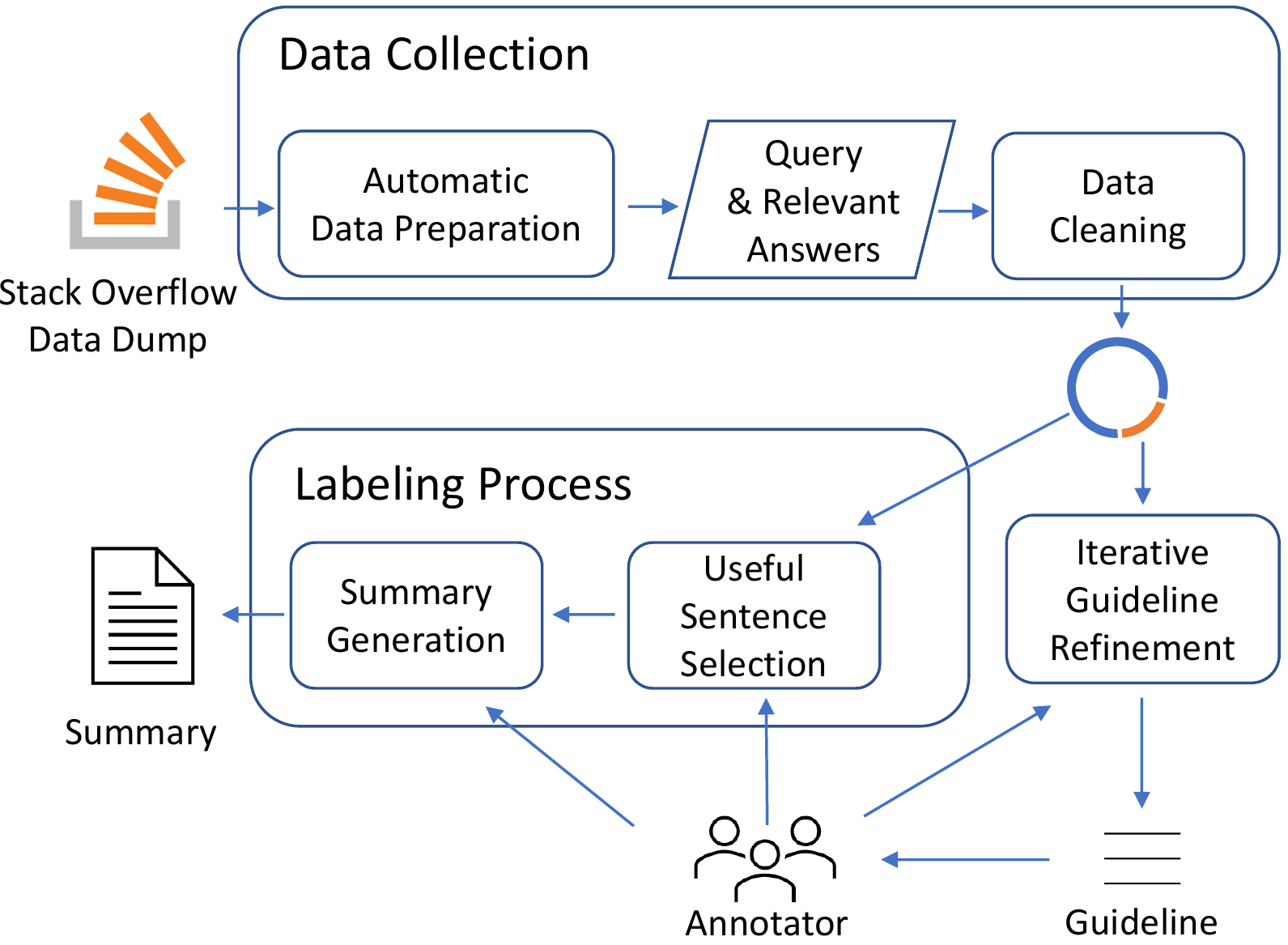}
    \vspace{-6mm}
    \caption{Benchmark Construction}
    \label{fig:bench_diagram}\vspace{-4mm}
\end{figure}

We present \bench{}, the first Stack Overflow multi-answer summarization benchmark for answering a specific technical question.
Figure~\ref{fig:bench_diagram} describes our benchmark construction process. We firstly automatically collect \emph{annotation units}, which is then followed by the data cleaning process described in Section~\ref{sec:datacollection}.
An annotation unit consists of a technical question and multiple relevant answers.
Next, six annotators perform a two-phase labeling process: useful sentence selection and summary generation as explained in Section~\ref{sec: labelingprocess}. 
Note that we perform an iterative guideline refinement for the useful sentence selection task until the inter-annotator agreement achieves a certain level as described in Section \ref{sec: useful sentence selection}. 


\subsection{Data Collection}\label{sec:datacollection}
We prepare 50 annotation units (i.e., technical query and multiple relevant answers) by performing automatic data extraction and data cleaning.

\subsubsection{\textbf{Automatic Data Preparation}}\label{automatic_data}

The PostLinks table in \so{} data dump\footnote{\url{https://archive.org/details/stackexchange}} contains the information of user-voted duplicated question links, in which each original question is linked to their duplicated questions.
Specifically, an original question refers to the earliest published \so{} question that is a duplicate of newer questions. The duplicate questions\footnote{https://stackoverflow.com/help/duplicates} are manually marked by following a strict mechanism that one moderator or at least three active users (i.e., users who have over 3,000 reputation score) vote the questions to be a duplicate to the original one. 
In particular, we treat the title of the original question as the technical question of each annotation unit in \bench{}. As the \so{} questions with `duplicate' links are labeled by reputable \so{} users, we regard all the answers to the original question and their duplicates as the relevant answers of the technical question in the corresponding annotation unit. 

To prepare the labeling materials, we randomly extract 50 annotation units from the PostLinks table in Stack Overflow data dump. 
We focus on questions related to two popular programming languages, Python and Java, by checking if they are tagged with `Java' or `Python'. 
Following AnswerBot~\cite{xu2017answerbot}, we only consider the technical questions in which the sum of the number of answers is between 10 to 15. We discard answers with no vote as their usefulness to the technical question is unclear or considered as bad by \so{} users. As we focus on the text summarization, answers that only contain code snippets without text content are also dropped.

\subsubsection{\textbf{Data Cleaning}}
We pre-process the 50 annotation units. We use NLTK\footnote{https://www.nltk.org/api/nltk.tokenize.html} to break each text content into sentences. 
We then apply some heuristics to reduce noise. 
We exclude sentences that consist of only an external hyperlink since the hyperlink content is not directly visible. 
For sentences that consist of both hyperlinks and text content, we keep the text content and replace the hyperlinks with a place holder (i.e., `[external-link]').  
We also replace the code snippet, table and pictures in answers with place holders (i.e., `[code-snippet]', `[table]', `[figure]'). Additionally, we observe that the inline code is important for annotators to understand the sentence meanings. Therefore, we keep the inline code. 
Furthermore, we treat multi-level headings which is more than five words in answers as individual sentences as we observe that they are frequently summative sentences.
Finally, each annotation unit after the above pre-processing serves as the labeling materials. 
\subsection{Labeling Process}
\label{sec: labelingprocess}

Firstly we prepare labeling materials for annotators to get ready for the labeling process.
We present the content of each annotation unit into a text file that contains the technical question, sentences from each answer, \so{} hyperlinks to each answer, and a blank section for labeling. Each answer sentence is assigned with a unique identifier based on its source answer and the sentence order in the source answer. For example, the fourth sentence in the third answer of the annotation unit is assigned with the identifier $\#03\_04$. We also provide the guideline.

Then we follow a two-phase labeling process that is commonly used in NLP studies \cite{angelidis2021extractive, angelidis2018summarizing}. 
Six annotators are involved in the labeling process.
All participants are PhD students and have at least two years of development experience in Java and Python programming languages.
In each phase, annotators and labeling materials are divided equally into two groups. Three participants in one group label the same group of labeling materials. 
The goal of the labeling process is to select summative sentences from multiple answers to form a brief summary of answers to a technical question.
To achieve this, in the first phase (i.e., \textit{useful sentence selection}), three annotators select the useful sentences from all relevant answers.
The sentences that all annotators agreed to be useful for answering the technical question serve as the input of the second phase (i.e., \textit{summary generation}).
In the \textit{summary generation} phase, three annotators are asked to select five sentences that will form an extractive summary based on the sentences' clarity, redundancy and importance. In prior summarization works from NLP field, it is common to select a fixed number of sentences~\cite{parveen2015topical, liu2019text, zhong2020extractive} , e.g., five sentences~\cite{parveen2015topical, zhong2020extractive}, as the summary.

\vspace{-3mm}
\subsubsection{\textbf{Phase 1: Useful Sentence Selection}}\label{sec: useful sentence selection}
In this labeling stage, annotators are required to identify the useful sentences for answering the technical question from multiple \so{} answers. 
The labeling objective is formulated as a binary classification that specifies sentence usefulness to the technical question. 
Annotators label the sentences for answering the technical question as either `useful' or `not-useful'. Sentences are labeled as `?' when annotators are unsure whether the sentences are useful to answer the question. Note that we do not consider the redundancy among sentences in this phase. 
In particular, we use the Kappa coefficient \cite{cohen1960coefficient} to measure the inter-annotator agreement. A higher kappa value illustrates a higher agreement level among annotators. We follow prior studies~\cite{joblin2017classifying, lee2017understanding,mantyla2015rapid} to interpret the meaning of Kappa value: kappa values 0 as poor agreement, 0.01–0.20 as slight agreement, 0.21–0.40 as fair, 0.41– 0.60 as moderate, 0.61–0.80 as substantial, and 0.81–1.00 as almost perfect agreement.



\noindent\textbf{Iterative Guideline Refinement} To ensure the quality of the selected useful sentences, we performed an iterative guideline refinement to improve the inter-annotator agreement, which is widely adopted in the literature~\cite{chang2017revolt, sheng2008get}. Specifically, six annotators label the same annotation units in each iteration. In each round, we polish the guideline and update the labeling materials according to the annotator feedback and automatic evaluation result. 
We end the guideline refinement process when the inter-annotator agreement level is \textit{Moderate}. 
The annotation units used in the guideline refinement process are discarded. Specifically, we discard 10 annotation units in the guideline refinement process
and have 40 annotation units for full labeling.

\noindent\textbf{Full Labeling Process} 
Firstly we arrange meetings with the annotators to align the understanding of the guideline. We share the labeling rules that are obtained while observing the guideline refinement process with annotators. In particular, annotators should identify the usefulness of each sentence independently of its context (e.g., surrounding sentences). 
Additionally, general comments about the question itself or its solutions should be avoided (e.g., ``I hate Java!''), as they do not provide useful information to answer the query. Finally, for incomplete sentences, annotators should not assume the usefulness of its hidden part (e.g., ``the content in `[code snippet]' place holder''). 

After the labeling process is finished, one author deals with the labeling results. In each annotation unit, all the sentences that three annotators agree to be useful to that question, coupled with the technical query, serve as the input to the second phase. 
Note that, there are three annotation units whose inter-annotator agreement is \textit{Slight} (i.e., the Kappa value $k$ is 0.00 $\leq$ $k$ $\geq$ 0.20) in the full labeling process. We consider a good agreement to be at least \textit{Moderate} (i.e., the Kappa value $k$ is 0.40 $\leq$ $k$ $\geq$ 0.60). Therefore, we discard these annotation units as the labels may not be of good quality. We have 37 sets of $\langle$ question, useful answers $\rangle$ pairs in total.  

\noindent\textbf{Inter-Annotator Agreement of Phase 1}
The average kappa value of the \textit{useful sentence selection} phase is 0.43, indicating the annotators achieve \textit{Moderate} agreement.
Previous work shows that the human agreements on sentence selection task for constructing news-domain summarization benchmarks are usually lower than 0.3 ~\cite{radev2003evaluation}. Additionally, the kappa value of a recent opinion summarization benchmark is 0.36~\cite{angelidis2021extractive}. It indicates that the inter-annotators agreement on sentence selection task in our \bench{} is comparable to those of common \nlp{} benchmarks. 

\vspace{-2mm}
\subsubsection{\textbf{Phase 2: Summary Generation}}

In the second phase, given a set of useful sentences to answer the technical question, annotators are required to generate a final extractive summary under a budget of five sentences by selecting sentences based on three key factors, i.e.,clarity of each sentence, redundancy among sentences, and the importance for each sentence to answer the question.

To let annotators equip a consensus of the goal, we arrange a meeting with all the annotators for two things.
First, we provide an example (excluded from the data used from constructing the benchmark) to demonstrate the materials and the brief overall labeling process.
Second, we let them align their understanding based on the materials by discussing with each other.

After that, annotators select sentences by following the process: first, all sentences are clustered into different RCs (Redundant Cluster) in which each sentence is semantically similar; then the annotator should select the most summative sentence with high clarity from each RC into the candidate list; finally the annotators are asked to delete sentences from the candidate list until only five sentences are left, by taking into account how well each sentence answers the question. 
Note that, for each group of sentences in our benchmark, there are always more than five RCs clustered by annotators.
Finally, the top five sentences serve as the summary of the answers to the technical question.

\vspace{-2mm}
\subsection{Benchmark Statistic}



In \bench{}, there are 37 technical questions, each of which corresponds to three ground truth answer summaries that are independently labeled by three annotators. Each summary contains five sentences extracted from corresponding relevant answers.
In total, we have 111 query-summary pairs. The average number of words in ground-truth summaries is 106.45. 
The scale of \bench{} is comparable with two commonly used query-focused multi-documentation summarization benchmarks in the NLP community~\cite{dang2005overview, gao2021simcse}, i.e., DUC-2005 and DUC-2007, which consist of 50 and 45 annotation units, respectively. 


\section{Approach}
\begin{figure*}[htbp]
    \centering
    \includegraphics[width=\textwidth]{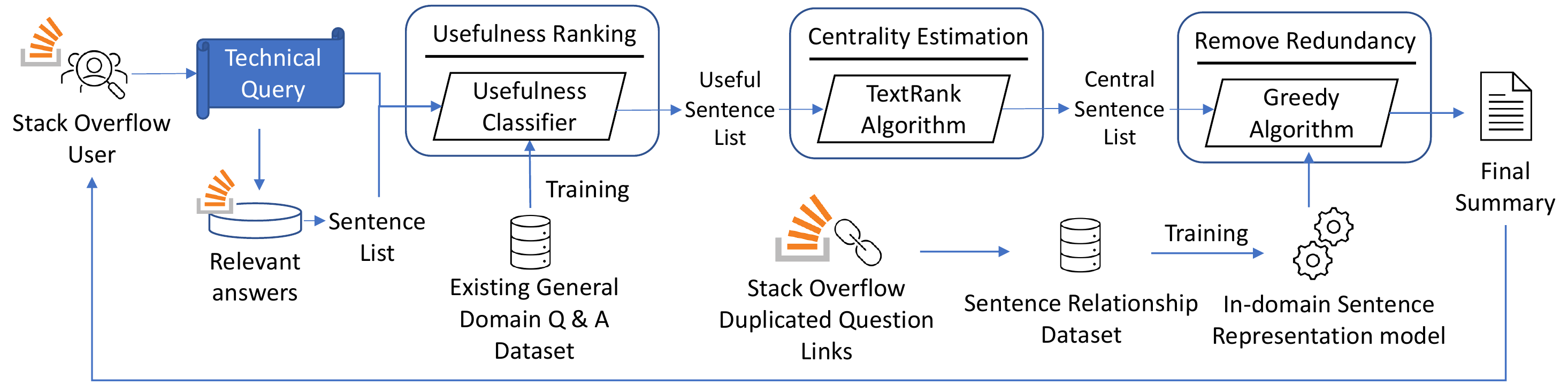}
    \vspace{-7mm}
    \caption{Overview of \toolname{}}
    \label{fig:end2end_diagram}
    \vspace{-5mm}
\end{figure*}

\label{sec:method}

This section presents our proposed approach \toolname{} to tackle the problem of answer summarization for technical queries. 
\vspace{-2mm}
\subsection{Overview}\label{sec:method overview}

As shown in Figure~\ref{fig:end2end_diagram}, \toolname{} takes a technical query and a set of relevant answers as input and outputs an extractive answer summary.
The framework of \toolname{} contains three core modules: \firstmodule{}, \secondmodule{}, and \thirdmodule{}.
Correspondingly, they tackle the problem from three different perspectives, 1) the degree of usefulness carried by an answer sentence to answer the query;
2) the centrality (importance) level carried by a sentence among answer sentences;
3) the extent of redundant information carried by two sentences, respectively.

\toolname{} firstly decomposes all the relevant answers into a list of sentences in the pre-processing step.
In the first \firstmodule{} module, we leverage a transfer learning-based approach to capture the \emph{usefulness} of each sentence to the given query.
\firstmodule{} module produces a ranked list of answer sentences ordered by the predicted usefulness.
Next, the second module, named \secondmodule{}, measures the centrality (i.e., the importance) of a sentence among the set of useful sentences by TextRank~\cite{mihalcea2004textrank}, a simple yet effective approach for the problem.
\secondmodule{} module outputs a rank list of sentences in terms of sentence centrality.
The third module \thirdmodule{} aims to reduce redundant sentences among candidates. To achieve the goal, we propose a greedy algorithm-based selection mechanism, in which we integrate the state-of-the-art sentence representation model.
The Redundancy Removal module also outputs a rank list of sentences without redundant sentences.
In the end, the top-5 answer sentences are used to form an answer summary to the target technical query.

\vspace{-2mm}
\subsection{Module I: Usefulness Ranking}\label{firstmodule}

Measuring the usefulness of sentences with respect to a given query is a common step for the technical query-focused answer summarization.
For example, AnswerBot~\cite{xu2017answerbot} assesses the usefulness by using handcrafted features which carry certain limitations. 
Taking one of the features named \emph{semantic patterns} as an example, it uses a static set with 12 manually summarized strings (e.g., ``\emph{I’d recommend}...'') as an indicator of usefulness.
However, such heuristic-based features require great human effort to maintain.

\subsubsection{\textbf{Transfer Learning-based Ranking Approach}}
Inspired by a recent work \cite{xu2020coarse}, we leverage BERT, a transfer learning-based model~\cite{devlin2018bert} that has achieved great success in many Stack Overflow-based tasks~\cite{zhang2020sentiment,yang2022aspect}, to model the extent of the usefulness of each answer sentence with respect to the query. Specifically, we first train a BERT-based classifier that predicts the likelihood of an answer sentence's usefulness score with respect to the target query. And then we rank the usefulness of each answer sentence by extracting the intermediate score (i.e., the probability of positive class) predicted by the classifier. 
We leverage a large-scale QA dataset named ASNQ~\cite{garg2020tanda} (short for Answer Sentence Natural Questions) from the general domain to fine-tune the BERT-based classifier. 
ASNQ has around 19 million pairs of data, in which each data is in the form of two sets of query-sentence pairs, i.e., $D = \{\{\langle Q_{i} \rightarrow PS_{j}\rangle \},\{\langle Q_{i} \rightarrow NS_{k}\rangle \}\}$. 
$Q_{i}$ denotes each query in ASNQ data, which is the real anonymized query issued to the Google search engine. $PS_{j}$ and $NS_{k}$ denote corresponding useful and useless Wikipedia sentence for answering the query.
Garg et al. \cite{garg2020tanda} argued that the BERT model trained with ASQN dataset can produce promising performance on NLP domain sentence usefulness classification tasks. By doing so, our model can benefit from the existing large-scale dataset.


\subsubsection{\textbf{Model Training}}\label{sec:ptm}
Firstly we train a BERT model for the sentence usefulness classification task by feeding the ASNQ dataset \cite{garg2020tanda}.
For each of the pairs $\langle Q,S\rangle$ in the ASNQ dataset, we convert it into a format acceptable by BERT, i.e., $\{[CLS],Q,[SEP],S,[SEP]\}$. Both $[CLS]$ and $[SEP]$ are special tokens used to form the input of BERT. $[CLS]$ stands for classification and $[SEP]$ stands for separator used to separate the $Q$ and $S$. We follow the standard padding (i.e., adding $[PAD]$ token) to the sequence to fix the length of each input instance to 512. Next, the $[CLS]$ vectors extracted from encoder output serves as input to the final classification layer (i.e., a single neural network layer) and Sigmoid function to produce the likelihood of the input instance to be predicted as positive or negative class. By following the standard practice, we use cross entropy as the loss function:

\vspace{-3mm}
\begin{equation}
    \mathcal{Loss}=-\sum_{i=1}^{N}{(y\log(p_{pos}) + (1 - y)\log p_{neg})}
\end{equation}

\noindent where $N$ denotes the number of training instances; $p_{pos}$ and $p_{neg}$ denotes the probability of the positive and negative classes. 

Once the BERT model is trained, for a given query $Q$ and a set of relevant answer sentences $S = \{S_1, S_2,..., S_i\}$, we utilize the model to predict the likelihood of each of the pairs $\langle Q, S_i\rangle$ as positive class (i.e., $P(class = useful | \langle Q, S\rangle)$) to the query and consider the probability $p_{pos}$ as the usefulness score. 
At the end, the usefulness score is used to rank all candidate sentences.
We select top $k$ answer sentences with the highest usefulness scores as the input of the next module.

\subsection{Module II: Centrality Estimation}\label{module2}

To select summative sentences, we consider the usefulness of each sentence to the query and the centrality of each sentence among all candidates.
To estimate the central sentences from all candidates, we apply a simple yet effective approach, TextRank \cite{mihalcea2004textrank}, to extract the representativeness score of each sentence. TextRank is a graph-based sentence ranking approach that essentially aims to quantify the \textit{representativeness} of each sentence in the sentence-graph based on global information recursively drawn from the entire graph. 
More precisely, a sentence similar to other sentences \textit{recommends} others and thus represents the overall understanding of the complete documentation. Meanwhile, a highly \textit{recommended} sentence by others is likely to be more summative and representative.

To perform TextRank algorithm, each candidate sentence is represented into a node in a sentence-graph with undirected edges. The initial weights of each edge are based on the word overlap between two nodes. 
Specifically, given two sentences $S_{i}$ and $S_{j}$, each sentence is in form of $N$ word tokens $t$, $S_{i}=t_{1}^{i},t_{2}^{i},...,t_{N}^{i}$, the initial edge weight is defined as: 

\vspace{-4mm}
\begin{equation}
Edge\_Weight(S_{i}, S_{j})=\frac{\left | \left \{ t_{k}|t_{k}\in S_{i} \& t_{k}\in S_{j}  \right \} \right |}{log\left ( \left | S_{i} \right | \right )+log\left ( \left | S_{j} \right | \right )}
\end{equation}

\noindent Then TextRank recursively calculates the representativeness score $R(s_{i})$ of each sentence $S_{i}$ as follow: 

\vspace{-3mm}
\begin{equation}
    R(S_{i})=1-\phi +\phi*\sum_{j\in In(S_{i})}^{}\frac{1}{Out(S_{i})}R(S_{j})
\end{equation}
\vspace{-2mm}

\noindent where $In(S_{i})$ denotes the list of nodes that point to $S_{i}$ while $Out(S_{i})$ denotes the list of nodes that $S_{i}$ points to. The $\phi\in(0,1)$ serves as the dumping factor; it refers to the probability of jumping from a given node to another random node in the graph (i.e., simulate the user behavior of web surfing in PageRank). We set the heuristic value $\phi$ as 0.85 which is the same as the original paper \cite{mihalcea2004textrank}. We randomly set the initial value of $R(s_{i})$ as the final value is not affected by the initial value $R(s_{i})$. TextRank stops the iterations when the convergence score $C$ is below a given threshold. Given the iteration $k$, the convergence score $C$ is defined as follows: 
\begin{equation}\vspace{-3mm}
    C = R^{k+1}(s_{i})-R^{k}(s_{i})
\end{equation}\vspace{-2mm}

\noindent As Mihalcea and Tarau \cite{mihalcea2004textrank} suggested, we set the threshold of $C$ as 0.0001. 
Finally, we rank all input sentences $s_{i}$ by taking the final representativeness score $R(s_{i})$ as the ranking score. We feed the ranked sentence list into the final module. 

\subsection{Module III: Redundancy Removal}\label{sec:redundancy}
According to the survey conducted by Xu et al.~\cite{xu2017answerbot}, redundancy is one of the key factors that developers face in identifying their target information in software questions and answer sites.
Thus, redundancy removal should be performed as a part of the answer summarization solution.
To remove redundant sentences, an in-domain sentence representation model is desired to measure the similarity between two answer sentences.
To achieve this, we create a large-scale SE domain sentence relationship dataset for training the sentence representation model. Each instance in this dataset is automatically extracted from Stack Overflow data by considering the implied semantics in the duplicate question links. We then use our collected dataset to train a contrastive learning-based sentence representation model and transform each sentence into the corresponding representation. 
Finally, we embed the learned representation into a greedy search algorithm to select sentences iteratively.



\subsubsection{\textbf{Dataset Construction for Contrastive Learning}}\label{sec:embedding_data}
To capture the semantic similarity between answer sentences, we train a Stack Overflow-specific sentence representation model based on contrastive learning.
The main goal of contrastive representation learning is to learn such an embedding space in which similar sentence pairs stay close to each other while dissimilar ones are far apart~\cite{gao2021simcse}.
Each input instance for training the approach is in the form of a triplet, $\langle s, s^{+}, s^{-}\rangle$, while $\langle s, s^{+}\rangle$ are considered as similar pairs and $\langle s, s^{-}\rangle$ are dissimilar ones.

Unfortunately, none of the existing datasets in SE domain directly serves the purpose. To tackle the problem, we first carefully observe the characteristic of Stack Overflow data and find that the duplicate question pairs (as described in Section \ref{automatic_data}) in Stack Overflow naturally carry the semantic relatedness between questions (i.e., describe the same semantic information but in different ways) by its definition.\footnote{\url{https://stackoverflow.com/help/duplicates}}
Duplicate questions could be a valuable signal considered as similar sentence pairs to supervise sentence representation, i.e., the representation of duplicate questions is expected to be close in the vector space while the non-duplicate pairs are far from each other.
Thus, we propose to create positive (i.e., similar) sentence pairs in the triplet by leveraging duplicate question pairs in Stack Overflow.
Specifically, we extract the titles of a pair of duplicate questions as the pair of the original sentence and its similar sentence, i.e., $\langle s, s^{+}\rangle$.
Besides, we also observe that if two questions are labeled by users with totally different tags, then they are most likely irrelevant.
Based on the observation, we randomly select pairs of questions that share no common tag and use their title to form the negative (i.e., dissimilar) sentence pairs (i.e., $\langle s, s^{-}\rangle$).
By following the \bench{} setting, we focus on Stack Overflow duplicate posts with `Java' and `Python' tags.
As a result, we built a dataset with 304,046 sentence triplets $\langle s, s^{+}, s^{-}\rangle$.

\subsubsection{\textbf{Contrastive Learning for Sentence Embedding}}\label{sec:contrastive_learning}
We apply SIMCSE~\cite{gao2021simcse}, the state-of-the-art sentence embedding approach based on contrastive learning structure.
SIMCSE uses a pre-trained model, RoBERTa \cite{liu2019roberta}, as its base model for generating embeddings and add a multilayer perceptron (i.e., MLP) layer on the top of it.

In the training stage of SIMCSE, for each input instance in the form of a triplet $\langle s,s ^{+}_{i}, s^{-}_{i}\rangle$ described in Section \ref{sec:embedding_data}, the loss function is defined as: 

\begin{equation}\label{equation:2}\vspace{-4mm}
    \mathcal{Loss}=-log\frac{e^{sim(r_{i},r_{i}^{+})/\tau}}{\sum_{j=1}^{N}(e^{sim(r_{i},r_{j}^{+})/\tau}+e^{sim(r_{i},r_{j}^{-})/\tau})}
\end{equation}

\noindent while $r_{i}$ denotes the representation of each input $s_{i}$, $N$ as the number of sentences in a mini-batch, and $\tau$ is a temperature hyperparameter, $sim(r_{1},r_{2})$ is the cosine similarity, respectively. Specifically, all parameters in base model RoBERTa \cite{liu2019roberta} are fine-tuned by minizing the loss function (i.e., Eq. \ref{equation:2}). We train the model by leveraging our in-domain dataset described in Section~\ref{sec:embedding_data}.


\subsubsection{\textbf{Greedy Algorithm for Redundancy Removal}}
The goal of this step is to select a subset of answer sentences with the minimum redundancy.
To achieve this, we apply a greedy algorithm which is originally proposed in the prior extractive summarization study R2N2 \cite{cao2015ranking} for removing redundant sentences.
Thus, it perfectly matches our scenario.

\vspace{-3mm}
\begin{algorithm}
\caption{Greedy Selection for Redundancy Removal}\label{alg:summarygeneration}
\SetKwComment{Comment}{/* }{ */}
\SetKwInput{KwInput}{Input}                
\KwInput{R[]: Ranked sentence list}
\KwResult{summary\_list: List of summary sentences}

embedding\_list = [] \Comment*[]{initialize empty embedding list}

\For{sent \textbf{in} R}{
    embed = generate\_embedding(sent)\;
    embedding\_list.append(embed)\;
    
}

redundancy\_removed = []\Comment*[]{initialize empty summary list}
redundancy\_removed.append(S[0])

\For{embed \textbf{in} embedding\_list[1:]}{
    \For{ground\_truth \textbf{in} redundancy\_removed}{

        \eIf{is\_redundant(embed, ground\_truth)}
            {continue}
        {redundancy\_removed.append(embed)}
    }
}

summary\_list = select\_top\_five(redundancy\_removed)

return summary\_list

\end{algorithm}
\vspace{-5mm}

In our case, we first transform all the input (i.e., answer sentences) into sentence embedding by using the sentence embedding model described in Section~\ref{sec:contrastive_learning} (i.e., Lines 2 to 5).
Then we pick each sentence for the final summary one-by-one according to the ranking given in the Centrality Estimation module in descending order (i.e., Line 8 to 16). For each selection, if the cosine similarity between the current sentence and others that are already included in the final summary is above a threshold $T$, the current sentence is regarded as redundant with the final summary and we discard it (i.e., Line 10 to 14).  
In the end, we pick the top five sentences after greedy selection to form the final summary (i.e., Line 17).
If less than five sentences are left after running the greedy selection algorithm, we pick all the remaining sentences as the summary.


\subsection{Implementation Details}

\subsubsection{\textbf{Usefulness Ranking Module}}\label{dataset1}

We implement the BERT model by using a popular Python deep learning library named \emph{Hugging Face}\footnote{\url{https://huggingface.com}}. The model version we select is `bert-base-uncased'. We tune the same key hyper-parameters (i.e., the learning rate, batch size, and number of epochs) as mentioned in the original BERT paper\cite{devlin2018bert}. We fine tune the model on the training set of ASNQ (around 19 million data instances) and tune the hyper-parameters (i.e., learning rate, batch size, and number of epochs) through the grid searching on the validation dataset of ASNQ.

\subsubsection{\textbf{Centrality Estimation Module}}\label{sec:data_for_module3}

We implement the TextRank algorithm by adapting the python TextRank library.\footnote{\url{https://github.com/summanlp/TextRank}}.
Unlike the original output of TextRank, which is in the form of a documentation, we extract the intermediate result of TextRank (i.e., the pairs of a sentence and its representativeness score) as the output.

\subsubsection{\textbf{Redundancy Removal Module}}
We split our in-domain sentence relationship dataset (described in Section~\ref{sec:embedding_data}) into training and test data with the ratio 9:1.
To train our in-domain sentence embedding model, we implement the base model (RoBERTa \cite{liu2019roberta}) by using the same python library \emph{Hugging Face} as Module I. 
We select the `roberta-base' verison.
We then reuse the RoBERTa model checkpoint provided by SIMCSE\footnote{https://github.com/princeton-nlp/SimCSE} that has been fine tuned with general domain dataset and further fine tune it with our training sub-dataset.
Meanwhile, we follow the same tuning methodology as tuning the BERT-based model for Usefulness Ranking Module (Section~\ref{dataset1}). We found that the performance achieved by tuning with different hyper-parameters does not differ much.
Thus we set the learning rate as 5e-5, batch size as 64, and the number of epochs as 3, which are the same as the setting in SIMCSE~\cite{gao2021simcse}.
We tune the threshold $T$ for redundancy removal by performing a duplicate SO post classification task only using titles, i.e., predict whether a given pair of Stack Overflow title sentences are duplicates or not. This is inspired by prior work~\cite{he2022ptm4tag} that the title of the Stack Overflow posts is the most informative component for expressing the semantics. Specifically, we obtain the representations of both sentences by using our trained sentence embedding model as described in Section~\ref{sec:contrastive_learning}. If the cosine similarity of both representations is higher than the threshold $T$, we consider them duplicates. We empirically investigate the performance with different values of the threshold and obtain the best performing value of 0.8.

\section{Experiment Settings}
This section describes the implementation details of all the baselines and the metrics that we used for evaluating the techniques automatically.

\subsection{Baselines}\label{baseline}
By following AnswerBot~\cite{xu2017answerbot}, we position our task as an \textit{extractive query-focused multi-documentation summarization} task.
Thus we compare \toolname{} against two groups of baselines: 1) the state-of-the-art answer summarization approach in the SE domain, AnswerBot~\cite{xu2017answerbot}; and 2) the state-of-the-art extractive query-focused and multi-doc summarization approaches in the NLP domain (i.e., LexRank \cite{erkan2004lexrank}, Biased-TextRank \cite{kazemi2020biased}, and QuerySum \cite{xu2020coarse}). 

\noindent\textbf{AnswerBot}. AnswerBot is the state-of-the-art Stack Overflow answer summarization approach. It retrieves relevant questions and then selects useful answer paragraphs. Finally, AnswerBot generates summaries by leveraging MRR algorithm. 
In relevant question retrieval process of AnswerBot \cite{xu2017answerbot}, it requires a relevant score, which represents the extent of the relevance of each answer for the query. In our experiment, we set the relevant scores as 1 because the input answers are selected from the duplicated question posts of the query, which is voted as duplicated by Stack Overflow users as described in Section \ref{benchmark}.

\noindent\textbf{LexRank}.
LexRank \cite{erkan2004lexrank} is a widely used text summarization approach that is based on sentence-graph. 
It computes the importance of sentences based on the cosine similarity in the sentence-graph. We implement the LexRank by leveraging python lexrank library.\footnote{https://pypi.org/project/lexrank/}

\noindent\textbf{Biased-TextRank}.
Biased-TextRank \cite{kazemi2020biased} achieves great performance on a query-focused debates summarization dataset in the NLP domain~\cite{kazemi2020biased}. It works for the query-focused multi-doc summarization task, which matches our task.
Biased-TextRank is an unsupervised summarization approach and based on the sentence graph. It also considers the query bias into the sentence-graph and applies an advanced sentence representation approach (i.e., Sentence-BERT \cite{reimers2019sentence}) in its pipeline. To adopt Biased-TextRank on \bench{}, we perform the grid search for its hyperparameters (i.e., damping\_factor and similarity\_score). 
We set both damping\_factor and similarity\_score to 0.7.

\noindent\textbf{QuerySum}.
To the best of our knowledge, QuerySum \cite{xu2020coarse} is the state-of-the-art summarization approach on query-focused multi-doc summarization task in the NLP domain. It follows coarse-to-fine framework that progressively estimates whether the sentences should be in the summary.
We modified the QuerySum \cite{xu2020coarse} pipeline to enable the sentence-level budget of the final summary other than the word number budget used in the original paper.

\subsection{Automatic Evaluation Metrics} \label{sec:eval_metrics}
Following the existing summarization approaches \cite{kazemi2020biased, xu2020coarse}, we use ROUGE (\textbf{R}ecall-\textbf{O}riented \textbf{U}nderstudy for \textbf{G}isting \textbf{E}valuation) \cite{lin2004rouge}, a set of evaluation metrics used for evaluationg the automatic summarization approaches, as our automatic evaluation metric.
ROUGE is a widely adopted evaluation metric for automatic evaluation for summarization systems \cite{fabbri2021summeval}.
We report ROUGE-N and ROUGE-L.
ROUGE-N evaluates the extent of the n-gram overlapping between a new summary and a set of ground-truth summaries \cite{lin2004rouge}:
\begin{equation}\vspace{-2mm}
\text{ROUGE-N}=\frac{\sum\limits_{S_{i}\in{S}}\sum\limits_{gram_{n}\in{S{{i}}}}Count_{match}(gram_{n})}{\sum\limits_{S_{i}\in{S}}\sum\limits_{gram_{n}\in{S{{i}}}}Count(gram_{n})}
\end{equation}
where $S$ denotes the set of ground-truth summaries, $n$ denotes the n-gram length of the summary, $Count_{match}(gram_{n})$ and  $gram_{n}$ denote the maximum n-grams numbers of their coexistence in the ground-truth summaries and a new summary. 
ROUGE-L measures the overlapping of LCS (i.e., longest common sub-sequence) between the new summary and ground-truth summaries. 
By following baseline~\cite{kazemi2020biased}, we report ROUGE-1, ROUGE-2, and ROUGE-L. We implement ROUGE evaluation by using the pyrouge python library.\footnote{https://github.com/bheinzerling/pyrouge}



\section{Experiment Results}
This section presents our experiment results with corresponding analysis to answer the following three research questions:

\vspace{0.2cm}

\noindent \textbf{RQ1}: How effective are the existing approaches in summarizing answers for technical queries on our benchmark? 

\noindent \textbf{RQ2}: Comparing with existing approaches, how effective is our approach \toolname{}?

\noindent \textbf{RQ3}: How much does each module contribute to the performance of \toolname{}?


\subsection{RQ1: Effectiveness of Existing Approaches}\label{empiricalstudy}

\noindent\textbf{Experimental setting.}
To answer this research question, we compare the performance of four existing approaches from both NLP and SE-domain as described in Section \ref{baseline}.
In particular, to investigate the improvement space of each approach, we also calculate the upper bound of the considered evaluation metrics by assuming the human-annotated summaries (i.e., ground truth) as the output of the ideal approach.
For each query in our benchmark, we calculate the ROUGE-N score (we set n = 1, 2) and ROUGE-L between the output produced by an approach and each of the ground truth summaries in our benchmark as described in Section~\ref{sec:eval_metrics}. Then, we compute the mean of the ROUGE scores.

\begin{table}[!htbp]
    \centering
    \vspace{-2mm}
    \caption{Overall Performance of Existing Approaches}
    \vspace{-3mm}
    \begin{tabular}{c|ccc}

    \toprule
    Approach & ROUGE-1  & ROUGE-2  & ROUGE-L     \\
    \midrule
    Upper Bound$^{*}$     &  0.784&  0.697 & 0.772\\
    \hline\multicolumn{4}{c}{Software Engineering Domain}\\\hline
    AnswerBot & 0.490& 0.276 & 0.456\\
    \hline\multicolumn{4}{c}{NLP Domain}\\\hline
    LexRank & 0.501& \textbf{0.289} & 0.448\\
    Biased-TextRank & 0.428& 0.217 & 0.400\\
    QuerySum & \textbf{0.508}&0.284&\textbf{0.476}\\
    \bottomrule
    \end{tabular}\\
    \small{
$^{*}$: Upper Bound: assuming the human-annotated summaries (i.e., ground truth summary) as the output of the ideal approach.}
\vspace{-3mm}
    \label{table: overallperformance}
\end{table}
\noindent\textbf{Result \& Analysis.}
As shown in the Table \ref{table: overallperformance}, we find that none of the considered approaches consistently performs better than the others.
Specifically, QuerySum performs the best in terms of ROUGE-1 and ROUGE-L while LexRank performs the best in terms of ROUGE-2.
More precisely, QuerySum and LexRank achieve comparable performance in terms of ROUGE-1 and ROUGE-2 but QuerySum outperforms LexRank on ROUGE-L by a large margin, i.e., 6\%.
AnswerBot performs worse than QuerySum and LexRank on all the evaluation metrics consistently but by a small margin.
Biased-TextRank performs the worst and its performance differs from the others.
Biased-TextRank and LexRank are unsupervised approaches and they are similar as both of them use the PageRank algorithm.
There are two key differences between them, (1) whether query information is considered, (2) integrated sentence representation techniques.
For the former difference, Biased-TextRank leverages query information (while LexRank does not) which has been demonstrated it is helpful in multiple related tasks, i.e., query-focus summarization.
It indicates that the integrated sentence embedding technique (i.e., Sentence-BERT) in Biased-TextRank could be its main bottleneck.
Overall, the performance ranking is QuerySum $>$ LexRank $>$ AnswerBot $>$ Biased-TextRank.

In particular, comparing the best performing approach from general domain (i.e., QuerySum) with domain-specific approach (i.e., AnswerBot), we find that QuerySum outperforms AnswerBot on all the evaluation metrics consistently but only by a small margin, i.e., 3.6\%, 2.8\%, 4.3\%, in terms of ROUGE-1, ROUGE-2, and ROUGE-L, respectively.

Comparing the existing approaches with the upper bound performance, we find that there is a big room for improvement.
The ROUGE-1, ROUGE-2, and ROUGE-L scores of best performing approach QuerySum is 25.29\%, 48.36\%, 28.21\% lower than the upper bound performance, respectively.
Particularly, we observe that improvement space in terms of ROUGE-2 is much bigger than that of ROUGE-1 (1.9 times) and ROUGE-L (1.7 times).
Considering ROUGE-2 refers to the bi-gram overlapping between generated summary and ground truth summary, it indicates that the existing approaches are unable to capture the information of domain-specific bi-gram terms well enough.





\begin{center}
\noindent\includegraphics{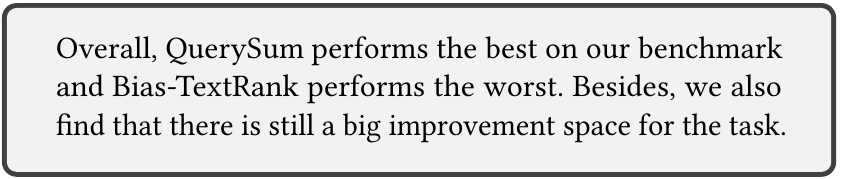}
\end{center}

\subsection{RQ2: Effectiveness of \toolname{}}\label{rq2}


\vspace{0.2cm}

\noindent\textbf{Experimental setting.}
To answer this research question, we firstly follow the same methodology of RQ1 and collect the performance (i.e., in terms of ROUGE-1, ROUGE-2, and ROUGE-L) of our approach on our benchmark.
In such a way, we can compare our proposed approach with the best performing baseline approaches from both software engineering (i.e., AnswerBot) and natural language processing (i.e., QuerySum) domains identified in RQ1.
Moreover, we conduct a user study on the three approaches to comprehensively understand their effectiveness.

The user study involves five participants, four of them are PhD students and the other one is a postdoctoral fellow. Note that none of participants are involved in the benchmark construction.
Considering the expensive human resources, we randomly sample a subset of our \bench{} with ten technical queries as the experimental data. Half of the queries are Python-related and the other half are Java-related.
All the participants have at least three years of development experience in Java or Python.
For each question, we collect three answer summaries generated by QuerySum, AnswerBot, and our approach, respectively. By following AnswerBot~\cite{xu2017answerbot}, we then ask all the participants to provide 5-point likert scores to all the summaries separately in terms of usefulness and diversity with respect to the technical query. In other words, participants are required to label each summary from ``extremely useless'' (denote as 1) to ``extremely useful'' (denote as 5) and from ``extremely redundant'' (denote as 1) to ``extremely diverse'' (denote as 5). For all the approaches, we report the average score labeled by participants across all the 10 queries as their final score. For each of the queries, we anonymize the corresponding approaches of all the summaries and randomly sort them to avoid bias.

\begin{table}
\vspace{-2mm}
    \centering
    \caption{Performance Comparison Between \toolname{} and Selected Approaches}
    \vspace{-3mm}
    \begin{tabular}{c|ccc}
    \toprule
    System & ROUGE-1 (\%)  & ROUGE-2 (\%) & ROUGE-L (\%)    \\
    \midrule
    \multicolumn{4}{c}{Software Engineering Domain}\\\hline
    AnswerBot & 0.490 (14.90) & 0.276 (36.59) & 0.456 (17.54) \\
    \hline
    \multicolumn{4}{c}{NLP Domain}\\\hline
    QuerySum & 0.508 (10.83) &0.284 (32.75) &0.476 (12.61)\\
    \hline
    \multicolumn{4}{c}{Our Approach}\\\hline
    \toolname{} & \textbf{0.563} & \textbf{0.377} & \textbf{0.536}\\
    \bottomrule
    \multicolumn{4}{c}{{\small \%: relative improvement of \toolname{}, i.e., (\toolname{}-}}\\
    \multicolumn{4}{c}{{\small baseline)/baseline.}}\\
    \end{tabular}
\vspace{-8mm}
    \label{table: ourperformance}
\end{table}

\noindent\textbf{Result \& Analysis.}
We report the automatic evaluation result in Table \ref{table: ourperformance}. 
The result shows that \toolname{} consistently outperforms best performing baseline approaches from both SE and NLP domains in terms of all evaluation metrics.
Comparing with the Software Engineering domain state-of-the-art (SOTA) baseline, i.e., AnswerBot, \toolname{} achieves better performance in terms of ROUGE-1, ROUGE-2, and ROUGE-L by 14.90\%, 36.59\%, and 17.54\%, respectively.
Comparing with QuerySum, the SOTA approach in \nlp, \toolname{} outperforms it by 10.83\%, 32.75\%, and 12.61\%, in terms of ROUGE-1, ROUGE-2, and ROUGE-L, respectively.
In summary, \toolname{} outperforms the existing approaches by a large margin.

\vspace{-4mm}
\begin{table}[!htbp]
    \centering
    \caption{User Study Result}
    \vspace{-3mm}
    \begin{tabular}{c|cc}

    \toprule
    System & Usefulness  & Diversity     \\
    \midrule
    AnswerBot & 3.68 & 3.62\\
    QuerySum & 3.80 & 3.64\\
    \toolname{} & \textbf{4.02}& \textbf{4.26}\\

    \bottomrule
    \end{tabular}\\

    \label{table: userstudy}\vspace{-3mm}
\end{table}
The result of the user study is shown in Table \ref{table: userstudy}. We report the mean of usefulness and diversity score of our approach and the two baselines. We observe that our approach also achieves the best performance consistently and the improvement is significant (p<0.05) in terms of both usefulness and diversity.
For the usefulness score, the average score of our approach is 4.02, i.e., between ``useful'' and ``extremely useful'', while AnswerBot and QuerySum are 3.68 and 3.80, i.e., between ``natural'' and ``useful''. In other words, our approach outperforms AnswerBot and QuerySum by 9.23\% and 5.79\% in terms of average usefulness score. For diversity score, the average score of our approach is 4.26, i.e., ``diverse'', while AnswerBot and QuerySum are 3.62 and 3.64, i.e., between ``neutral'' and ``diverse''. Our approach outperforms AnswerBot and QuerySum by 17.68\% and 17.03\% in terms of average diversity score. Overall, both automatic evaluation and user study demonstrate that our approach outperforms the existing approaches.
Meanwhile, we observe that Answerbot's diversity and usefulness scores in our user study are very close to those in the user study of Answerbot~\cite{xu2017answerbot}.
We also find that the result derived from automatic evaluation based on our benchmark and human evaluation are consistent. Both of them indicate that our approach produces the most promising answer summaries.



\begin{center}
\noindent\includegraphics{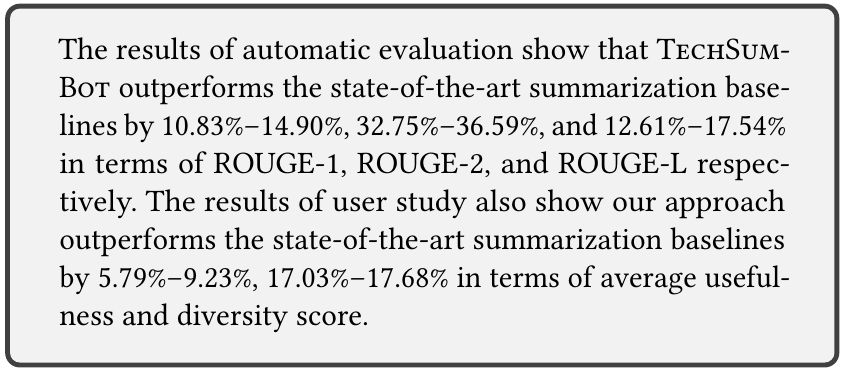}
\end{center}


\subsection{RQ3: Ablation Study}\label{ablation}


\noindent\textbf{Experimental setting.}
We conduct an ablation study on the performance of each module following the same setting for evaluating QuerySum in \cite{xu2020coarse}.
Since each module outputs a rank list of answer sentences, we iteratively integrate them one by one and collect the top-5 ranked sentences as the final summary.
Then we follow the same setting of evaluating the end-to-end approach and calculate the evaluation metrics, i.e., ROUGE-1, ROUGE-2, and ROUGE-L.

\begin{table}
\vspace{-2mm}
    \centering
    \caption{Ablation Study of Each Module}
    \vspace{-3mm}
    \begin{tabular}{c|ccc} 
    \toprule
    System & ROUGE-1  & ROUGE-2  & ROUGE-L     \\
    \midrule
    \multicolumn{4}{c}{\toolname{}}\\\hline
    Module I & 0.507& 0.298 & 0.478\\
    Module I \& II & 0.543 & 0.340 & 0.512 \\
    Module I \& II \& III & \textbf{0.563} & \textbf{0.377} & \textbf{0.536} \\
    \bottomrule
    \multicolumn{4}{c}{\small{Module I: Usefulness Ranking Module. Module II: Centrality Estimation }}\\
    \multicolumn{4}{c}{\small{Module. Module III: Redundancy Removal Module.}}
    \end{tabular}
    \label{table: ablationstudy}
    \vspace{-5mm}
\end{table}

\vspace{0.2cm}
\noindent\textbf{Result \& Analysis.}
We report the result in Table \ref{table: ablationstudy}.
We observe that each module contributes its own part to the system performance.
By only applying \textit{Module I}, we observe that it can already consistently outperforms AnswerBot in terms of all evaluation metrics by a small margin.
Besides, \textit{Module I} can achieve comparable result with QuerySum. The result provides another evidence on the effectiveness of pre-trained model.
Comparing with \textit{Module I}, integrating \textit{Module II} on the top of \textit{Module I} can further boost the performance by 7.10\%, 17.11\%, and 7.11\% in terms of ROUGE-1, ROUGE-2, and ROUGE-L, respectively.
Besides, we find that it can already achieve better performance than all the existing approaches in terms of all the evaluation metrics consistently.
By combining all the modules, i.e., the complete version of our approach, it produces the best performance. Specifically, it improves the performance over applying the first two modules by 3.68\%, 8.02\%, and 4.69\% in terms of ROUGE-1, ROUGE-2, and ROUGE-L respectively.

\begin{center}
\noindent\includegraphics{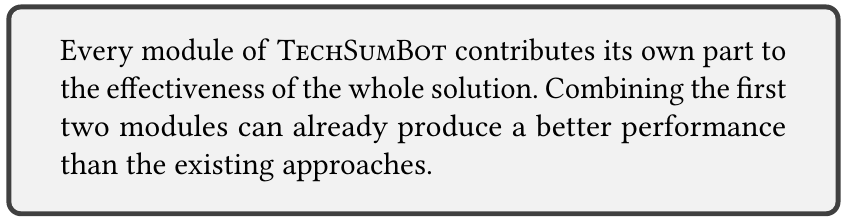}
\end{center}

\section{Discussion}

\subsection{Qualitative Analysis}\label{sec:qualitative}
We conduct a qualitative analysis by comparing the summaries generated by each baseline with the ground-truth summaries in \bench{}. 
More precisely, we analyze the sentences in ground-truth summaries that are incorrectly discarded by baseline approaches and sentences that appear in the summaries generated by baselines but excluded in the ground-truth summaries.
We identified multiple weaknesses of existing approaches. 

First, we observe that the semantic patterns used in AnswerBot sometimes lead to incorrect results. 
For example, 
given a query in \bench{}: 
``\textit{What is the difference between a primary key and a unique constraint}'',
there is one relevant answer sentence\footnote{https://stackoverflow.com/a/9391610} that annotators do not consider useful:
``\textit{I suggest that if you are the kind of coder, ..., then PRIMARY KEY is just the thing for you}''.
\toolname{} successfully determines that this sentence is not useful to the query.
However, a heuristic semantic pattern in AnswerBot regards sentences that start with ``\textit{I suggest that ...}'' as useful, which leads AnswerBot to incorrectly identify the sentence to be useful. 
As curating high-quality patterns for extractive summarization task relies on the deep understanding of the dataset characteristics \cite{zhong2019closer}, pattern-based summarization approaches on \so{} may requires in-depth exploration of Stack Overflow data.


Second, we observe that existing approaches are weak at distinguishing the redundant sentences between answer sentences. For example, annotators consider both sentences ``2**2 = 2 power-of 2.'' and ``The ** operator in Python is really "power;" that is, 2**3 = 8.'' useful to the query, but redundant to each other since both sentences explain `**' operator in Python. To avoid the redundancy in the summary, only one of the two sentences should be retained in the final summary. However, QuerySum and AnswerBot put both sentences into their final summaries. The root cause of such a failure is that the two queries are semantically close but lexically different. When removing redundant sentences, both QuerySum and AnswerBot represent sentences by bag-of-words model and word embedding, which cannot capture the semantic information. 

\vspace{-2mm}
\subsection{How Automatic Evaluation Complements User Evaluation}

Though user evaluation has advantages~\cite{buse2011benefits}, researchers have perceived some barriers when performing user evaluations~\cite{buse2011benefits, le2019reliability, fabbri2021summeval}. Especially, recruiting issues and time commitment are the two most common barriers\cite{buse2011benefits}. Fabbri et al.~\cite{fabbri2021summeval} also report that existing summarization approaches apply user evaluations that are largely inconsistent with previous works. Furthermore, they observe that the annotation judges of evaluating summarization approaches are different among annotators with dissimilar knowledge backgrounds (e.g., experts and crowd-sourced workers)~\cite{fabbri2021summeval}.

In our case, for query-focused Stack Overflow answer summarization, an automatic evaluation approach by using our benchmark \bench{} can mitigate the aforementioned barriers. By offering a ground-truth summarization benchmark, automatically evaluating a summarization approach no longer requires recruiting annotators and long waiting periods. Besides, the automatic evaluation result is reusable for the subsequent works by applying the same evaluation metric, e.g., ROUGE~\cite{lin2004rouge} on the same benchmark.

We acknowledge that the current version of our benchmark may still contain subjective bias due to the limited size of our benchmark and the annotators' background and judgment preferences, which are also the biases that exist when conducting a user study. We mitigate such a bias by offering multiple ground-truth summaries that are produced by different annotators independently for the same query. The automatic evaluation's result is calculated by using the averaged values of the considered evaluation metrics. The biggest advantage of our benchmark is that it's highly extendable as compared with human evaluation. We provide the replication package as well as the guideline so that the benchmark can be further expanded by community effort. 
For example, researchers can independently provide more summaries for each query or expand the pool of the queries. With such contributions from the community, the subjective bias can be continuously mitigated.

\subsection{Threats to Validity}
\noindent Threats to internal validity are related to the implementation errors of \toolname{} and baselines. To mitigate the first threat, we have double checked our code. For the implementation of baseline approaches, the threat is limited as we reuse their published replication packages. The experiment bias of the annotators may also affect the internal validity. To minimize the threat, we conduct an iterative guideline refinement process.
In addition, all the annotators have Java or Python development experience for at least three years.
\noindent Threats to external validity are related to the generalizability of our benchmark and experiment results.
To ensure the quality of the benchmark, we follow the standard process to produce summaries and updated the guidelines iteratively. 
The programming language considered in our experiments is also a threat to external validity. 
Different from prior work~\cite{xu2017answerbot} that only considers Stack Overflow posts tagged as Java, our work mitigates this threat by considering two popular programming languages, i.e., Java and Python. 

\noindent Threats to construct validity are related to the used evaluation metrics. ROUGE is widely used as an automatic evaluation metric for summarization approaches in both NLP and SE domains~\cite{kazemi2020biased,xu2020coarse,zhang2022summary}. We use usefulness and diversity in our human evaluation, which are widely used to evaluate the SE domain summarization tasks~\cite{mani2012ausum,usefulness2016, xu2017answerbot}. Thus, we believe the threat is minimal.

\section{Related Work}

Summarization of software artifacts has gained much research interest.
There are multiple previous works on summarizing different contents in Stack Overflow. Xu et al.~\cite{xu2017answerbot} proposed AnswerBot which is a Stack Overflow answer summary generator with respect to a specific query.
Besides, Opiner~\cite{uddin2017automatic} aims to help developers efficiently understand API by summarizing API reviews.
Opiner adopts available summarization algorithms, such as Textrank~\cite{mihalcea2004textrank} (extractive) and Opinosis~\cite{ganesan2010opinosis} (abstractive), to produce API reviews from Stack Overflow posts.
Furthermore, Nadi and Treude~\cite{nadi2020essential} extracted the essential sentences from Stack Overflow to navigate the developers reading answers.
Naghshzan et al.~\cite{naghshzan2021leveraging} proposed an approach based on TextRank algorithm to summarize Android API methods discussed in Stack Overflow.
Their approach is based on extractive summarization that extracts the most important sentences from the Stack Overflow posts.
The differences between prior works and ours are three-fold. First, different from the aforementioned works that extract useful Stack Overflow sentences by using handcrafted features, we achieve the goal by proposing a transfer learning-based approach. We train the pre-trained models to learn pre-knowledge from a large-scale QA dataset. Second, previous approaches remove redundant sentences in a simple way which carries certain limitations. For example, to calculate sentence similarity, AnswerBot represents sentences into vectors through word embedding and word IDF metrics, which cannot capture the semantic-related sequential information and is poor at distinguishing sentences that are semantically different but lexically similar~\cite{wei2022clear}. Differently, \toolname{} combines both transformer-based sentence representation model and contrastive learning training approach to address the limitations. Third, unlike aforementioned works that perform manual evaluation like user studies, we construct the first benchmark for SO query-focused multi-answer summarization task and enable automatic evaluation, which carries unique advantages as compared with human evaluation.

Apart from automatic summarization for Stack Overflow, NLP community researchers also work on summarization tasks, e.g., query-focused summarization and multi-documentation summarization.
Xu et al.~\cite{xu2020coarse} presents that the obstacles to multi-doc summarization over the setting of single-doc are (1) difficulty in obtaining training data, (2) large size and number of input resources, and (3) redundancy among input resources.
Traditional summarization approaches (e.g., LexRank~\cite{erkan2004lexrank}) rely on the sentence graph and apply PageRank algorithm to rank the sentences.
Biased-TextRank~\cite{kazemi2020biased} is query-focused. It considers embedding the query bias into the graph and leveraging Sentence-BERT~\cite{zheng2019sentence} to represent the sentences. QuerySum, to our best knowledge, is the state-of-the-art query-focused and multi-doc summarization approach in NLP field. It follows the `coarse-to-fine' principle to select and filter sentences. QuerySum also considers the relationship between queries and sentences as a useful sentence selection task.
Different from the above mentioned works, our approach's framework is specifically designed for Stack Overflow answer summarization. It takes into account the question-answer data structure as well as the widespread content redundancy in software Q\&A sites~\cite{xu2017answerbot}. Besides, our approach leverages the characteristic of Stack Overflow data. For instance, we observed that Stack Overflow users actively label the duplication relationship between posts. Hence, we leverage the information to train the in-domain sentence representation model.


\vspace{-1mm}
\section{Conclusion and Future Work}
In this paper, we focus on the problem of Stack Overflow answer summarization for technical queries.
Considering the limitations of only conducting human evaluation without automatic evaluation in the past studies, we find that there is a need of a benchmark to enable automatic evaluation as a complementary evaluation method.
Thus, we manually construct the first \so{} multi-answer summarization benchmark \bench{}, which consists of 111 query-answer summary pairs from 382 Stack Overflow answers with 2,014 sentence candidates. 
Based on \bench{}, we evaluate the performance of existing and potential applicable approaches from both SE and NLP fields. The results indicate the approach from the NLP field (i.e., QuerySum~\cite{xu2020coarse}) achieves the best performance and there is still a big room for improvement. Motivated by this, we further propose a new approach \toolname{} to tackle the problem. We perform both automatic and human evaluations to evaluate \toolname{} and the results show that \toolname{} outperforms all SOTA baselines from both SE and NLP fields by a large margin.  
In the future, we plan to integrate our approach \toolname{} into an IDE to help developers in searching their needed information from SQA sites more efficiently and accurately.
\vspace{1mm}

\begin{acks}
This research / project is supported by the Ministry of Education, Singapore, under its Academic Research Fund Tier 2 (Award No.: MOE2019-T2-1-193). Any opinions, findings and conclusions or recommendations expressed in this material are those of the author(s) and do not reflect the views of the Ministry of Education, Singapore.
\end{acks}

\balance
\bibliographystyle{ACM-Reference-Format}
\bibliography{ref}


\begin{thebibliography}{48}


\ifx \showCODEN    \undefined \def \showCODEN     #1{\unskip}     \fi
\ifx \showDOI      \undefined \def \showDOI       #1{#1}\fi
\ifx \showISBNx    \undefined \def \showISBNx     #1{\unskip}     \fi
\ifx \showISBNxiii \undefined \def \showISBNxiii  #1{\unskip}     \fi
\ifx \showISSN     \undefined \def \showISSN      #1{\unskip}     \fi
\ifx \showLCCN     \undefined \def \showLCCN      #1{\unskip}     \fi
\ifx \shownote     \undefined \def \shownote      #1{#1}          \fi
\ifx \showarticletitle \undefined \def \showarticletitle #1{#1}   \fi
\ifx \showURL      \undefined \def \showURL       {\relax}        \fi
\providecommand\bibfield[2]{#2}
\providecommand\bibinfo[2]{#2}
\providecommand\natexlab[1]{#1}
\providecommand\showeprint[2][]{arXiv:#2}

\bibitem[\protect\citeauthoryear{Angelidis, Amplayo, Suhara, Wang, and
  Lapata}{Angelidis et~al\mbox{.}}{2021}]%
        {angelidis2021extractive}
\bibfield{author}{\bibinfo{person}{Stefanos Angelidis},
  \bibinfo{person}{Reinald~Kim Amplayo}, \bibinfo{person}{Yoshihiko Suhara},
  \bibinfo{person}{Xiaolan Wang}, {and} \bibinfo{person}{Mirella Lapata}.}
  \bibinfo{year}{2021}\natexlab{}.
\newblock \showarticletitle{Extractive opinion summarization in quantized
  transformer spaces}.
\newblock \bibinfo{journal}{\emph{Transactions of the Association for
  Computational Linguistics}}  \bibinfo{volume}{9} (\bibinfo{year}{2021}),
  \bibinfo{pages}{277--293}.
\newblock


\bibitem[\protect\citeauthoryear{Angelidis and Lapata}{Angelidis and
  Lapata}{2018}]%
        {angelidis2018summarizing}
\bibfield{author}{\bibinfo{person}{Stefanos Angelidis} {and}
  \bibinfo{person}{Maria Lapata}.} \bibinfo{year}{2018}\natexlab{}.
\newblock \showarticletitle{Summarizing Opinions: Aspect Extraction Meets
  Sentiment Prediction and They Are Both Weakly Supervised}. In
  \bibinfo{booktitle}{\emph{2018 Conference on Empirical Methods in Natural
  Language Processing}}. Association for Computational Linguistics,
  \bibinfo{pages}{3675--3686}.
\newblock


\bibitem[\protect\citeauthoryear{Buse, Sadowski, and Weimer}{Buse
  et~al\mbox{.}}{2011}]%
        {buse2011benefits}
\bibfield{author}{\bibinfo{person}{Raymond~PL Buse}, \bibinfo{person}{Caitlin
  Sadowski}, {and} \bibinfo{person}{Westley Weimer}.}
  \bibinfo{year}{2011}\natexlab{}.
\newblock \showarticletitle{Benefits and barriers of user evaluation in
  software engineering research}. In \bibinfo{booktitle}{\emph{Proceedings of
  the 2011 ACM international conference on Object oriented programming systems
  languages and applications}}. \bibinfo{pages}{643--656}.
\newblock


\bibitem[\protect\citeauthoryear{Cao, Wei, Dong, Li, and Zhou}{Cao
  et~al\mbox{.}}{2015}]%
        {cao2015ranking}
\bibfield{author}{\bibinfo{person}{Ziqiang Cao}, \bibinfo{person}{Furu Wei},
  \bibinfo{person}{Li Dong}, \bibinfo{person}{Sujian Li}, {and}
  \bibinfo{person}{Ming Zhou}.} \bibinfo{year}{2015}\natexlab{}.
\newblock \showarticletitle{Ranking with recursive neural networks and its
  application to multi-document summarization}. In
  \bibinfo{booktitle}{\emph{Proceedings of the AAAI Conference on Artificial
  Intelligence}}, Vol.~\bibinfo{volume}{29}.
\newblock


\bibitem[\protect\citeauthoryear{Chang, Amershi, and Kamar}{Chang
  et~al\mbox{.}}{2017}]%
        {chang2017revolt}
\bibfield{author}{\bibinfo{person}{Joseph~Chee Chang}, \bibinfo{person}{Saleema
  Amershi}, {and} \bibinfo{person}{Ece Kamar}.}
  \bibinfo{year}{2017}\natexlab{}.
\newblock \showarticletitle{Revolt: Collaborative crowdsourcing for labeling
  machine learning datasets}. In \bibinfo{booktitle}{\emph{Proceedings of the
  2017 CHI Conference on Human Factors in Computing Systems}}.
  \bibinfo{pages}{2334--2346}.
\newblock


\bibitem[\protect\citeauthoryear{Cohen}{Cohen}{1960}]%
        {cohen1960coefficient}
\bibfield{author}{\bibinfo{person}{Jacob Cohen}.}
  \bibinfo{year}{1960}\natexlab{}.
\newblock \showarticletitle{A coefficient of agreement for nominal scales}.
\newblock \bibinfo{journal}{\emph{Educational and psychological measurement}}
  \bibinfo{volume}{20}, \bibinfo{number}{1} (\bibinfo{year}{1960}),
  \bibinfo{pages}{37--46}.
\newblock


\bibitem[\protect\citeauthoryear{Dang}{Dang}{2005}]%
        {dang2005overview}
\bibfield{author}{\bibinfo{person}{Hoa~Trang Dang}.}
  \bibinfo{year}{2005}\natexlab{}.
\newblock \showarticletitle{Overview of DUC 2005}. In
  \bibinfo{booktitle}{\emph{Proceedings of the document understanding
  conference}}, Vol.~\bibinfo{volume}{2005}. \bibinfo{pages}{1--12}.
\newblock


\bibitem[\protect\citeauthoryear{Di~Sorbo, Panichella, Alexandru, Shimagaki,
  Visaggio, Canfora, and Gall}{Di~Sorbo et~al\mbox{.}}{2016}]%
        {usefulness2016}
\bibfield{author}{\bibinfo{person}{Andrea Di~Sorbo},
  \bibinfo{person}{Sebastiano Panichella}, \bibinfo{person}{Carol~V.
  Alexandru}, \bibinfo{person}{Junji Shimagaki}, \bibinfo{person}{Corrado~A.
  Visaggio}, \bibinfo{person}{Gerardo Canfora}, {and}
  \bibinfo{person}{Harald~C. Gall}.} \bibinfo{year}{2016}\natexlab{}.
\newblock \showarticletitle{What Would Users Change in My App? Summarizing App
  Reviews for Recommending Software Changes}. In
  \bibinfo{booktitle}{\emph{Proceedings of the 2016 24th ACM SIGSOFT
  International Symposium on Foundations of Software Engineering}}
  \emph{(\bibinfo{series}{FSE 2016})}. \bibinfo{pages}{499–510}.
\newblock


\bibitem[\protect\citeauthoryear{Erkan and Radev}{Erkan and Radev}{2004}]%
        {erkan2004lexrank}
\bibfield{author}{\bibinfo{person}{G{\"u}nes Erkan} {and}
  \bibinfo{person}{Dragomir~R Radev}.} \bibinfo{year}{2004}\natexlab{}.
\newblock \showarticletitle{Lexrank: Graph-based lexical centrality as salience
  in text summarization}.
\newblock \bibinfo{journal}{\emph{Journal of artificial intelligence research}}
   \bibinfo{volume}{22} (\bibinfo{year}{2004}), \bibinfo{pages}{457--479}.
\newblock


\bibitem[\protect\citeauthoryear{Fabbri, Kry{\'s}ci{\'n}ski, McCann, Xiong,
  Socher, and Radev}{Fabbri et~al\mbox{.}}{2021}]%
        {fabbri2021summeval}
\bibfield{author}{\bibinfo{person}{Alexander~R Fabbri},
  \bibinfo{person}{Wojciech Kry{\'s}ci{\'n}ski}, \bibinfo{person}{Bryan
  McCann}, \bibinfo{person}{Caiming Xiong}, \bibinfo{person}{Richard Socher},
  {and} \bibinfo{person}{Dragomir Radev}.} \bibinfo{year}{2021}\natexlab{}.
\newblock \showarticletitle{Summeval: Re-evaluating summarization evaluation}.
\newblock \bibinfo{journal}{\emph{Transactions of the Association for
  Computational Linguistics}}  \bibinfo{volume}{9} (\bibinfo{year}{2021}),
  \bibinfo{pages}{391--409}.
\newblock


\bibitem[\protect\citeauthoryear{Ganesan, Zhai, and Han}{Ganesan
  et~al\mbox{.}}{2010}]%
        {ganesan2010opinosis}
\bibfield{author}{\bibinfo{person}{Kavita Ganesan}, \bibinfo{person}{ChengXiang
  Zhai}, {and} \bibinfo{person}{Jiawei Han}.} \bibinfo{year}{2010}\natexlab{}.
\newblock \showarticletitle{Opinosis: A graph based approach to abstractive
  summarization of highly redundant opinions}.
\newblock  (\bibinfo{year}{2010}).
\newblock


\bibitem[\protect\citeauthoryear{Gao, Yao, and Chen}{Gao et~al\mbox{.}}{2021}]%
        {gao2021simcse}
\bibfield{author}{\bibinfo{person}{Tianyu Gao}, \bibinfo{person}{Xingcheng
  Yao}, {and} \bibinfo{person}{Danqi Chen}.} \bibinfo{year}{2021}\natexlab{}.
\newblock \showarticletitle{SimCSE: Simple Contrastive Learning of Sentence
  Embeddings}. In \bibinfo{booktitle}{\emph{Proceedings of the 2021 Conference
  on Empirical Methods in Natural Language Processing}}.
  \bibinfo{pages}{6894--6910}.
\newblock


\bibitem[\protect\citeauthoryear{Garg, Vu, and Moschitti}{Garg
  et~al\mbox{.}}{2020}]%
        {garg2020tanda}
\bibfield{author}{\bibinfo{person}{Siddhant Garg}, \bibinfo{person}{Thuy Vu},
  {and} \bibinfo{person}{Alessandro Moschitti}.}
  \bibinfo{year}{2020}\natexlab{}.
\newblock \showarticletitle{Tanda: Transfer and adapt pre-trained transformer
  models for answer sentence selection}. In
  \bibinfo{booktitle}{\emph{Proceedings of the AAAI Conference on Artificial
  Intelligence}}, Vol.~\bibinfo{volume}{34}. \bibinfo{pages}{7780--7788}.
\newblock


\bibitem[\protect\citeauthoryear{He, Xu, Yang, Han, Yang, and Lo}{He
  et~al\mbox{.}}{2022}]%
        {he2022ptm4tag}
\bibfield{author}{\bibinfo{person}{Junda He}, \bibinfo{person}{Bowen Xu},
  \bibinfo{person}{Zhou Yang}, \bibinfo{person}{DongGyun Han},
  \bibinfo{person}{Chengran Yang}, {and} \bibinfo{person}{David Lo}.}
  \bibinfo{year}{2022}\natexlab{}.
\newblock \showarticletitle{PTM4Tag: Sharpening Tag Recommendation of Stack
  Overflow Posts with Pre-trained Models}.
\newblock  (\bibinfo{year}{2022}).
\newblock


\bibitem[\protect\citeauthoryear{Jacob, Ming-Wei, Kenton, and Kristina}{Jacob
  et~al\mbox{.}}{2019}]%
        {devlin2018bert}
\bibfield{author}{\bibinfo{person}{Devlin Jacob}, \bibinfo{person}{Chang
  Ming-Wei}, \bibinfo{person}{Lee Kenton}, {and} \bibinfo{person}{Toutanova
  Kristina}.} \bibinfo{year}{2019}\natexlab{}.
\newblock \showarticletitle{BERT: Pre-training of Deep Bidirectional
  Transformers for Language Understanding}. In
  \bibinfo{booktitle}{\emph{Proceedings of NAACL-HLT}}.
  \bibinfo{pages}{4171--4186}.
\newblock


\bibitem[\protect\citeauthoryear{Joblin, Apel, Hunsen, and Mauerer}{Joblin
  et~al\mbox{.}}{2017}]%
        {joblin2017classifying}
\bibfield{author}{\bibinfo{person}{Mitchell Joblin}, \bibinfo{person}{Sven
  Apel}, \bibinfo{person}{Claus Hunsen}, {and} \bibinfo{person}{Wolfgang
  Mauerer}.} \bibinfo{year}{2017}\natexlab{}.
\newblock \showarticletitle{Classifying developers into core and peripheral: An
  empirical study on count and network metrics}. In
  \bibinfo{booktitle}{\emph{2017 IEEE/ACM 39th International Conference on
  Software Engineering (ICSE)}}. IEEE, \bibinfo{pages}{164--174}.
\newblock


\bibitem[\protect\citeauthoryear{Kazemi, P{\'e}rez-Rosas, and Mihalcea}{Kazemi
  et~al\mbox{.}}{2020}]%
        {kazemi2020biased}
\bibfield{author}{\bibinfo{person}{Ashkan Kazemi},
  \bibinfo{person}{Ver{\'o}nica P{\'e}rez-Rosas}, {and} \bibinfo{person}{Rada
  Mihalcea}.} \bibinfo{year}{2020}\natexlab{}.
\newblock \showarticletitle{Biased TextRank: Unsupervised Graph-Based Content
  Extraction}. In \bibinfo{booktitle}{\emph{Proceedings of the 28th
  International Conference on Computational Linguistics}}.
  \bibinfo{pages}{1642--1652}.
\newblock


\bibitem[\protect\citeauthoryear{Le, Bao, Lo, Xia, Li, and Pasareanu}{Le
  et~al\mbox{.}}{2019}]%
        {le2019reliability}
\bibfield{author}{\bibinfo{person}{Xuan-Bach~D Le}, \bibinfo{person}{Lingfeng
  Bao}, \bibinfo{person}{David Lo}, \bibinfo{person}{Xin Xia},
  \bibinfo{person}{Shanping Li}, {and} \bibinfo{person}{Corina Pasareanu}.}
  \bibinfo{year}{2019}\natexlab{}.
\newblock \showarticletitle{On reliability of patch correctness assessment}. In
  \bibinfo{booktitle}{\emph{2019 IEEE/ACM 41st International Conference on
  Software Engineering (ICSE)}}. IEEE, \bibinfo{pages}{524--535}.
\newblock


\bibitem[\protect\citeauthoryear{Lee, Carver, and Bosu}{Lee
  et~al\mbox{.}}{2017}]%
        {lee2017understanding}
\bibfield{author}{\bibinfo{person}{Amanda Lee}, \bibinfo{person}{Jeffrey~C
  Carver}, {and} \bibinfo{person}{Amiangshu Bosu}.}
  \bibinfo{year}{2017}\natexlab{}.
\newblock \showarticletitle{Understanding the impressions, motivations, and
  barriers of one time code contributors to FLOSS projects: a survey}. In
  \bibinfo{booktitle}{\emph{2017 IEEE/ACM 39th International Conference on
  Software Engineering (ICSE)}}. IEEE, \bibinfo{pages}{187--197}.
\newblock


\bibitem[\protect\citeauthoryear{Lin}{Lin}{2004}]%
        {lin2004rouge}
\bibfield{author}{\bibinfo{person}{Chin-Yew Lin}.}
  \bibinfo{year}{2004}\natexlab{}.
\newblock \showarticletitle{Rouge: A package for automatic evaluation of
  summaries}. In \bibinfo{booktitle}{\emph{Text summarization branches out}}.
  \bibinfo{pages}{74--81}.
\newblock


\bibitem[\protect\citeauthoryear{Liu, Baltes, Treude, Lo, Zhang, and Xia}{Liu
  et~al\mbox{.}}{2021}]%
        {liu2021characterizing}
\bibfield{author}{\bibinfo{person}{Jiakun Liu}, \bibinfo{person}{Sebastian
  Baltes}, \bibinfo{person}{Christoph Treude}, \bibinfo{person}{David Lo},
  \bibinfo{person}{Yun Zhang}, {and} \bibinfo{person}{Xin Xia}.}
  \bibinfo{year}{2021}\natexlab{}.
\newblock \showarticletitle{Characterizing search activities on stack
  overflow}. In \bibinfo{booktitle}{\emph{Proceedings of the 29th ACM Joint
  Meeting on European Software Engineering Conference and Symposium on the
  Foundations of Software Engineering}}. \bibinfo{pages}{919--931}.
\newblock


\bibitem[\protect\citeauthoryear{Liu and Lapata}{Liu and Lapata}{2019}]%
        {liu2019text}
\bibfield{author}{\bibinfo{person}{Yang Liu} {and} \bibinfo{person}{Mirella
  Lapata}.} \bibinfo{year}{2019}\natexlab{}.
\newblock \showarticletitle{Text Summarization with Pretrained Encoders}. In
  \bibinfo{booktitle}{\emph{Proceedings of the 2019 Conference on Empirical
  Methods in Natural Language Processing and the 9th International Joint
  Conference on Natural Language Processing (EMNLP-IJCNLP)}}.
  \bibinfo{pages}{3730--3740}.
\newblock


\bibitem[\protect\citeauthoryear{Liu, Ott, Goyal, Du, Joshi, Chen, Levy, Lewis,
  Zettlemoyer, and Stoyanov}{Liu et~al\mbox{.}}{2019}]%
        {liu2019roberta}
\bibfield{author}{\bibinfo{person}{Yinhan Liu}, \bibinfo{person}{Myle Ott},
  \bibinfo{person}{Naman Goyal}, \bibinfo{person}{Jingfei Du},
  \bibinfo{person}{Mandar Joshi}, \bibinfo{person}{Danqi Chen},
  \bibinfo{person}{Omer Levy}, \bibinfo{person}{Mike Lewis},
  \bibinfo{person}{Luke Zettlemoyer}, {and} \bibinfo{person}{Veselin
  Stoyanov}.} \bibinfo{year}{2019}\natexlab{}.
\newblock \showarticletitle{Roberta: A robustly optimized bert pretraining
  approach}.
\newblock \bibinfo{journal}{\emph{arXiv preprint arXiv:1907.11692}}
  (\bibinfo{year}{2019}).
\newblock


\bibitem[\protect\citeauthoryear{Mahajan, Abolhassani, and Prasad}{Mahajan
  et~al\mbox{.}}{2020}]%
        {mahajan2020recommending}
\bibfield{author}{\bibinfo{person}{Sonal Mahajan}, \bibinfo{person}{Negarsadat
  Abolhassani}, {and} \bibinfo{person}{Mukul~R Prasad}.}
  \bibinfo{year}{2020}\natexlab{}.
\newblock \showarticletitle{Recommending stack overflow posts for fixing
  runtime exceptions using failure scenario matching}. In
  \bibinfo{booktitle}{\emph{Proceedings of the 28th ACM Joint Meeting on
  European Software Engineering Conference and Symposium on the Foundations of
  Software Engineering}}. \bibinfo{pages}{1052--1064}.
\newblock


\bibitem[\protect\citeauthoryear{Mani, Catherine, Sinha, and Dubey}{Mani
  et~al\mbox{.}}{2012}]%
        {mani2012ausum}
\bibfield{author}{\bibinfo{person}{Senthil Mani}, \bibinfo{person}{Rose
  Catherine}, \bibinfo{person}{Vibha~Singhal Sinha}, {and}
  \bibinfo{person}{Avinava Dubey}.} \bibinfo{year}{2012}\natexlab{}.
\newblock \showarticletitle{Ausum: approach for unsupervised bug report
  summarization}. In \bibinfo{booktitle}{\emph{Proceedings of the ACM SIGSOFT
  20th International Symposium on the Foundations of Software Engineering}}.
  \bibinfo{pages}{1--11}.
\newblock


\bibitem[\protect\citeauthoryear{M{\"a}ntyl{\"a}, Adams, Khomh, Engstr{\"o}m,
  and Petersen}{M{\"a}ntyl{\"a} et~al\mbox{.}}{2015}]%
        {mantyla2015rapid}
\bibfield{author}{\bibinfo{person}{Mika~V M{\"a}ntyl{\"a}},
  \bibinfo{person}{Bram Adams}, \bibinfo{person}{Foutse Khomh},
  \bibinfo{person}{Emelie Engstr{\"o}m}, {and} \bibinfo{person}{Kai Petersen}.}
  \bibinfo{year}{2015}\natexlab{}.
\newblock \showarticletitle{On rapid releases and software testing: a case
  study and a semi-systematic literature review}.
\newblock \bibinfo{journal}{\emph{Empirical Software Engineering}}
  \bibinfo{volume}{20}, \bibinfo{number}{5} (\bibinfo{year}{2015}),
  \bibinfo{pages}{1384--1425}.
\newblock


\bibitem[\protect\citeauthoryear{Mihalcea and Tarau}{Mihalcea and
  Tarau}{2004}]%
        {mihalcea2004textrank}
\bibfield{author}{\bibinfo{person}{Rada Mihalcea} {and} \bibinfo{person}{Paul
  Tarau}.} \bibinfo{year}{2004}\natexlab{}.
\newblock \showarticletitle{Textrank: Bringing order into text}. In
  \bibinfo{booktitle}{\emph{Proceedings of the 2004 conference on empirical
  methods in natural language processing}}. \bibinfo{pages}{404--411}.
\newblock


\bibitem[\protect\citeauthoryear{Nadi and Treude}{Nadi and Treude}{2020}]%
        {nadi2020essential}
\bibfield{author}{\bibinfo{person}{Sarah Nadi} {and} \bibinfo{person}{Christoph
  Treude}.} \bibinfo{year}{2020}\natexlab{}.
\newblock \showarticletitle{Essential sentences for navigating Stack Overflow
  answers}. In \bibinfo{booktitle}{\emph{2020 IEEE 27th International
  Conference on Software Analysis, Evolution and Reengineering (SANER)}}. IEEE,
  \bibinfo{pages}{229--239}.
\newblock


\bibitem[\protect\citeauthoryear{Naghshzan, Guerrouj, and Baysal}{Naghshzan
  et~al\mbox{.}}{2021}]%
        {naghshzan2021leveraging}
\bibfield{author}{\bibinfo{person}{AmirHossein Naghshzan},
  \bibinfo{person}{Latifa Guerrouj}, {and} \bibinfo{person}{Olga Baysal}.}
  \bibinfo{year}{2021}\natexlab{}.
\newblock \showarticletitle{Leveraging Unsupervised Learning to Summarize APIs
  Discussed in Stack Overflow}. In \bibinfo{booktitle}{\emph{2021 IEEE 21st
  International Working Conference on Source Code Analysis and Manipulation
  (SCAM)}}. IEEE, \bibinfo{pages}{142--152}.
\newblock


\bibitem[\protect\citeauthoryear{Nallapati, Zhou, Gulcehre, Xiang,
  et~al\mbox{.}}{Nallapati et~al\mbox{.}}{2016}]%
        {nallapati2016abstractive}
\bibfield{author}{\bibinfo{person}{Ramesh Nallapati}, \bibinfo{person}{Bowen
  Zhou}, \bibinfo{person}{Caglar Gulcehre}, \bibinfo{person}{Bing Xiang},
  {et~al\mbox{.}}} \bibinfo{year}{2016}\natexlab{}.
\newblock \showarticletitle{Abstractive Text Summarization using
  Sequence-to-sequence RNNs and Beyond}. In
  \bibinfo{booktitle}{\emph{Proceedings of The 20th SIGNLL Conference on
  Computational Natural Language Learning}}. \bibinfo{pages}{280--290}.
\newblock


\bibitem[\protect\citeauthoryear{Napoles, Gormley, and Van~Durme}{Napoles
  et~al\mbox{.}}{2012}]%
        {napoles2012annotated}
\bibfield{author}{\bibinfo{person}{Courtney Napoles},
  \bibinfo{person}{Matthew~R Gormley}, {and} \bibinfo{person}{Benjamin
  Van~Durme}.} \bibinfo{year}{2012}\natexlab{}.
\newblock \showarticletitle{Annotated gigaword}. In
  \bibinfo{booktitle}{\emph{Proceedings of the Joint Workshop on Automatic
  Knowledge Base Construction and Web-scale Knowledge Extraction
  (AKBC-WEKEX)}}. \bibinfo{pages}{95--100}.
\newblock


\bibitem[\protect\citeauthoryear{Parveen, Ramsl, and Strube}{Parveen
  et~al\mbox{.}}{2015}]%
        {parveen2015topical}
\bibfield{author}{\bibinfo{person}{Daraksha Parveen},
  \bibinfo{person}{Hans-Martin Ramsl}, {and} \bibinfo{person}{Michael Strube}.}
  \bibinfo{year}{2015}\natexlab{}.
\newblock \showarticletitle{Topical coherence for graph-based extractive
  summarization}. In \bibinfo{booktitle}{\emph{Proceedings of the 2015
  conference on empirical methods in natural language processing}}.
  \bibinfo{pages}{1949--1954}.
\newblock


\bibitem[\protect\citeauthoryear{Radev, Teufel, Saggion, Lam, Blitzer, Qi,
  Celebi, Liu, and Drabek}{Radev et~al\mbox{.}}{2003}]%
        {radev2003evaluation}
\bibfield{author}{\bibinfo{person}{Dragomir Radev}, \bibinfo{person}{Simone
  Teufel}, \bibinfo{person}{Horacio Saggion}, \bibinfo{person}{Wai Lam},
  \bibinfo{person}{John Blitzer}, \bibinfo{person}{Hong Qi},
  \bibinfo{person}{Arda Celebi}, \bibinfo{person}{Danyu Liu}, {and}
  \bibinfo{person}{Elliott~Franco Drabek}.} \bibinfo{year}{2003}\natexlab{}.
\newblock \showarticletitle{Evaluation challenges in large-scale document
  summarization}. In \bibinfo{booktitle}{\emph{Proceedings of the 41st Annual
  Meeting of the Association for Computational Linguistics}}.
  \bibinfo{pages}{375--382}.
\newblock


\bibitem[\protect\citeauthoryear{Reimers, Gurevych, Reimers, Gurevych, Thakur,
  Reimers, Daxenberger, Gurevych, Reimers, Gurevych, et~al\mbox{.}}{Reimers
  et~al\mbox{.}}{2019}]%
        {reimers2019sentence}
\bibfield{author}{\bibinfo{person}{Nils Reimers}, \bibinfo{person}{Iryna
  Gurevych}, \bibinfo{person}{Nils Reimers}, \bibinfo{person}{Iryna Gurevych},
  \bibinfo{person}{Nandan Thakur}, \bibinfo{person}{Nils Reimers},
  \bibinfo{person}{Johannes Daxenberger}, \bibinfo{person}{Iryna Gurevych},
  \bibinfo{person}{Nils Reimers}, \bibinfo{person}{Iryna Gurevych},
  {et~al\mbox{.}}} \bibinfo{year}{2019}\natexlab{}.
\newblock \showarticletitle{Sentence-BERT: Sentence Embeddings using Siamese
  BERT-Networks}. In \bibinfo{booktitle}{\emph{Proceedings of the 2019
  Conference on Empirical Methods in Natural Language Processing}}. Association
  for Computational Linguistics, \bibinfo{pages}{671--688}.
\newblock


\bibitem[\protect\citeauthoryear{Sheng, Provost, and Ipeirotis}{Sheng
  et~al\mbox{.}}{2008}]%
        {sheng2008get}
\bibfield{author}{\bibinfo{person}{Victor~S Sheng}, \bibinfo{person}{Foster
  Provost}, {and} \bibinfo{person}{Panagiotis~G Ipeirotis}.}
  \bibinfo{year}{2008}\natexlab{}.
\newblock \showarticletitle{Get another label? improving data quality and data
  mining using multiple, noisy labelers}. In
  \bibinfo{booktitle}{\emph{Proceedings of the 14th ACM SIGKDD international
  conference on Knowledge discovery and data mining}}.
  \bibinfo{pages}{614--622}.
\newblock


\bibitem[\protect\citeauthoryear{Silva, Roy, Rahman, Schneider, Paixao, and
  de~Almeida~Maia}{Silva et~al\mbox{.}}{2019}]%
        {silva2019recommending}
\bibfield{author}{\bibinfo{person}{Rodrigo~FG Silva},
  \bibinfo{person}{Chanchal~K Roy}, \bibinfo{person}{Mohammad~Masudur Rahman},
  \bibinfo{person}{Kevin~A Schneider}, \bibinfo{person}{Klerisson Paixao},
  {and} \bibinfo{person}{Marcelo de Almeida~Maia}.}
  \bibinfo{year}{2019}\natexlab{}.
\newblock \showarticletitle{Recommending comprehensive solutions for
  programming tasks by mining crowd knowledge}. In
  \bibinfo{booktitle}{\emph{2019 IEEE/ACM 27th International Conference on
  Program Comprehension (ICPC)}}. IEEE, \bibinfo{pages}{358--368}.
\newblock


\bibitem[\protect\citeauthoryear{Treude and Robillard}{Treude and
  Robillard}{2016}]%
        {treude2016augmenting}
\bibfield{author}{\bibinfo{person}{Christoph Treude} {and}
  \bibinfo{person}{Martin~P Robillard}.} \bibinfo{year}{2016}\natexlab{}.
\newblock \showarticletitle{Augmenting api documentation with insights from
  stack overflow}. In \bibinfo{booktitle}{\emph{2016 IEEE/ACM 38th
  International Conference on Software Engineering (ICSE)}}. IEEE,
  \bibinfo{pages}{392--403}.
\newblock


\bibitem[\protect\citeauthoryear{Uddin and Khomh}{Uddin and Khomh}{2017}]%
        {uddin2017automatic}
\bibfield{author}{\bibinfo{person}{Gias Uddin} {and} \bibinfo{person}{Foutse
  Khomh}.} \bibinfo{year}{2017}\natexlab{}.
\newblock \showarticletitle{Automatic summarization of API reviews}. In
  \bibinfo{booktitle}{\emph{2017 32nd IEEE/ACM International Conference on
  Automated Software Engineering (ASE)}}. IEEE, \bibinfo{pages}{159--170}.
\newblock


\bibitem[\protect\citeauthoryear{Wang, Phan, Wang, and Zhao}{Wang
  et~al\mbox{.}}{2019}]%
        {wang2019extracting}
\bibfield{author}{\bibinfo{person}{Shaohua Wang}, \bibinfo{person}{NhatHai
  Phan}, \bibinfo{person}{Yan Wang}, {and} \bibinfo{person}{Yong Zhao}.}
  \bibinfo{year}{2019}\natexlab{}.
\newblock \showarticletitle{Extracting API tips from developer question and
  answer websites}. In \bibinfo{booktitle}{\emph{2019 IEEE/ACM 16th
  International Conference on Mining Software Repositories (MSR)}}. IEEE,
  \bibinfo{pages}{321--332}.
\newblock


\bibitem[\protect\citeauthoryear{Wei, Harzevili, Huang, Wang, and Wang}{Wei
  et~al\mbox{.}}{2022}]%
        {wei2022clear}
\bibfield{author}{\bibinfo{person}{Moshi Wei}, \bibinfo{person}{Nima~Shiri
  Harzevili}, \bibinfo{person}{Yuchao Huang}, \bibinfo{person}{Junjie Wang},
  {and} \bibinfo{person}{Song Wang}.} \bibinfo{year}{2022}\natexlab{}.
\newblock \showarticletitle{CLEAR: Contrastive Learning for API
  Recommendation}. In \bibinfo{booktitle}{\emph{2022 IEEE/ACM 44th
  International Conference on Software Engineering (ICSE)}}. IEEE,
  \bibinfo{pages}{376--387}.
\newblock


\bibitem[\protect\citeauthoryear{Xu, Xing, Xia, and Lo}{Xu
  et~al\mbox{.}}{2017}]%
        {xu2017answerbot}
\bibfield{author}{\bibinfo{person}{Bowen Xu}, \bibinfo{person}{Zhenchang Xing},
  \bibinfo{person}{Xin Xia}, {and} \bibinfo{person}{David Lo}.}
  \bibinfo{year}{2017}\natexlab{}.
\newblock \showarticletitle{AnswerBot: Automated generation of answer summary
  to developers' technical questions}. In \bibinfo{booktitle}{\emph{2017 32nd
  IEEE/ACM International Conference on Automated Software Engineering (ASE)}}.
  IEEE, \bibinfo{pages}{706--716}.
\newblock


\bibitem[\protect\citeauthoryear{Xu and Lapata}{Xu and Lapata}{2020}]%
        {xu2020coarse}
\bibfield{author}{\bibinfo{person}{Yumo Xu} {and} \bibinfo{person}{Mirella
  Lapata}.} \bibinfo{year}{2020}\natexlab{}.
\newblock \showarticletitle{Coarse-to-fine query focused multi-document
  summarization}. In \bibinfo{booktitle}{\emph{Proceedings of the 2020
  Conference on empirical methods in natural language processing (EMNLP)}}.
  \bibinfo{pages}{3632--3645}.
\newblock


\bibitem[\protect\citeauthoryear{Yang, Xu, Khan, Uddin, Han, Yang, and Lo}{Yang
  et~al\mbox{.}}{2022}]%
        {yang2022aspect}
\bibfield{author}{\bibinfo{person}{Chengran Yang}, \bibinfo{person}{Bowen Xu},
  \bibinfo{person}{Junaed~Younus Khan}, \bibinfo{person}{Gias Uddin},
  \bibinfo{person}{Donggyun Han}, \bibinfo{person}{Zhou Yang}, {and}
  \bibinfo{person}{David Lo}.} \bibinfo{year}{2022}\natexlab{}.
\newblock \showarticletitle{Aspect-Based API Review Classification: How Far Can
  Pre-Trained Transformer Model Go?}. In \bibinfo{booktitle}{\emph{2022 IEEE
  International Conference on Software Analysis, Evolution and Reengineering
  (SANER). IEEE Computer Society}}.
\newblock


\bibitem[\protect\citeauthoryear{Zhang, Irsan, Thung, Han, Lo, and Jiang}{Zhang
  et~al\mbox{.}}{2022}]%
        {zhang2022summary}
\bibfield{author}{\bibinfo{person}{Ting Zhang}, \bibinfo{person}{Ivana~Clairine
  Irsan}, \bibinfo{person}{Ferdian Thung}, \bibinfo{person}{DongGyun Han},
  \bibinfo{person}{David Lo}, {and} \bibinfo{person}{Lingxiao Jiang}.}
  \bibinfo{year}{2022}\natexlab{}.
\newblock \showarticletitle{Automatic pull request title generation}. In
  \bibinfo{booktitle}{\emph{2022 IEEE International Conference on Software
  Maintenance and Evolution (ICSME)}}. IEEE.
\newblock


\bibitem[\protect\citeauthoryear{Zhang, Xu, Thung, Haryono, Lo, and
  Jiang}{Zhang et~al\mbox{.}}{2020}]%
        {zhang2020sentiment}
\bibfield{author}{\bibinfo{person}{Ting Zhang}, \bibinfo{person}{Bowen Xu},
  \bibinfo{person}{Ferdian Thung}, \bibinfo{person}{Stefanus~Agus Haryono},
  \bibinfo{person}{David Lo}, {and} \bibinfo{person}{Lingxiao Jiang}.}
  \bibinfo{year}{2020}\natexlab{}.
\newblock \showarticletitle{Sentiment analysis for software engineering: How
  far can pre-trained transformer models go?}. In
  \bibinfo{booktitle}{\emph{2020 IEEE International Conference on Software
  Maintenance and Evolution (ICSME)}}. IEEE, \bibinfo{pages}{70--80}.
\newblock


\bibitem[\protect\citeauthoryear{Zheng and Lapata}{Zheng and Lapata}{2019}]%
        {zheng2019sentence}
\bibfield{author}{\bibinfo{person}{Hao Zheng} {and} \bibinfo{person}{Mirella
  Lapata}.} \bibinfo{year}{2019}\natexlab{}.
\newblock \showarticletitle{Sentence Centrality Revisited for Unsupervised
  Summarization}. In \bibinfo{booktitle}{\emph{Proceedings of the 57th Annual
  Meeting of the Association for Computational Linguistics}}.
  \bibinfo{pages}{6236--6247}.
\newblock


\bibitem[\protect\citeauthoryear{Zhong, Liu, Chen, Wang, Qiu, and Huang}{Zhong
  et~al\mbox{.}}{2020}]%
        {zhong2020extractive}
\bibfield{author}{\bibinfo{person}{Ming Zhong}, \bibinfo{person}{Pengfei Liu},
  \bibinfo{person}{Yiran Chen}, \bibinfo{person}{Danqing Wang},
  \bibinfo{person}{Xipeng Qiu}, {and} \bibinfo{person}{Xuan-Jing Huang}.}
  \bibinfo{year}{2020}\natexlab{}.
\newblock \showarticletitle{Extractive Summarization as Text Matching}. In
  \bibinfo{booktitle}{\emph{Proceedings of the 58th Annual Meeting of the
  Association for Computational Linguistics}}. \bibinfo{pages}{6197--6208}.
\newblock


\bibitem[\protect\citeauthoryear{Zhong, Wang, Liu, Qiu, and Huang}{Zhong
  et~al\mbox{.}}{2019}]%
        {zhong2019closer}
\bibfield{author}{\bibinfo{person}{Ming Zhong}, \bibinfo{person}{Danqing Wang},
  \bibinfo{person}{Pengfei Liu}, \bibinfo{person}{Xipeng Qiu}, {and}
  \bibinfo{person}{Xuanjing Huang}.} \bibinfo{year}{2019}\natexlab{}.
\newblock \showarticletitle{A closer look at data bias in neural extractive
  summarization models}.
\newblock \bibinfo{journal}{\emph{arXiv preprint arXiv:1909.13705}}
  (\bibinfo{year}{2019}).
\newblock


\end{thebibliography}

\end{document}